\documentclass[reprint,
superscriptaddress,
nobibnotes,
amsmath,amssymb,
aps,
]{revtex4-2}

\usepackage[utf8]{inputenc} 
\usepackage[T1]{fontenc}    
\usepackage{hyperref}       
\usepackage{url}            
\usepackage{amsfonts}       
\usepackage{amsmath,amsthm,amssymb} 
\usepackage{bm}
\usepackage{bbm}
\usepackage{dcolumn}
\usepackage{nicefrac}       
\usepackage{graphicx}
\usepackage{xcolor, etoolbox}

\usepackage{mathtools}
\usepackage{mathrsfs}

\usepackage{ragged2e}
\usepackage{microtype}      
\usepackage{braket}      
\usepackage{enumitem}
\usepackage{xspace}
\usepackage{dsfont}

\usepackage{float}

\DeclarePairedDelimiter\abs{\lvert}{\rvert}

\hypersetup{
    colorlinks,
    linkcolor=blue,
    citecolor=blue,
    urlcolor=blue,
    breaklinks=true 
}

\begin{document}
\title{Scalar quantum fields in cosmologies with $2+1$ spacetime dimensions}

\author{Natalia Sánchez-Kuntz}
    \email{sanchez@thphys.uni-heidelberg.de}
    \affiliation{Institut f\"{u}r Theoretische Physik, Universit\"{a}t Heidelberg, \\ Philosophenweg 16, 69120 Heidelberg, Germany}
\author{Álvaro Parra-López}
    \email{alvaparr@ucm.es}
    \affiliation{Institut f\"{u}r Theoretische Physik, Universit\"{a}t Heidelberg, \\ Philosophenweg 16, 69120 Heidelberg, Germany}
    \affiliation{Departamento de F\'isica Te\'orica and IPARCOS, Facultad de Ciencias Físicas, Universidad Complutense de Madrid, Ciudad Universitaria, 28040 Madrid, Spain}
\author{Mireia Tolosa-Simeón}
    \email{mtolosa@thp.uni-koeln.de}
    \affiliation{Institut f\"{u}r Theoretische Physik, Universit\"{a}t Heidelberg, \\ Philosophenweg 16, 69120 Heidelberg, Germany}
\author{Tobias Haas}
    \email{t.haas@thphys.uni-heidelberg.de}
    \affiliation{Institut f\"{u}r Theoretische Physik, Universit\"{a}t Heidelberg, \\ Philosophenweg 16, 69120 Heidelberg, Germany} 
\author{Stefan Floerchinger}
    \email{stefan.floerchinger@uni-jena.de}
    \affiliation{Institut f\"{u}r Theoretische Physik, Universit\"{a}t Heidelberg, \\ Philosophenweg 16, 69120 Heidelberg, Germany}
    \affiliation{Theoretisch-Physikalisches Institut, Friedrich-Schiller-Universit\"{a}t Jena,\\
    Max-Wien-Platz 1, 07743 Jena, Germany}

\date{\today}

\begin{abstract}
Motivated by the possibility to use Bose-Einstein condensates as quantum simulators for spacetime curvature, we study a massless relativistic scalar quantum field in spatially curved Friedmann-Lema\^itre-Robertson-Walker universes with $d=2+1$ spacetime dimensions. In particular, we investigate particle production caused by a time-dependent background geometry, by means of the spectrum of fluctuations and several two-point field correlation functions. We derive new analytical results for several expansion scenarios.
\end{abstract}

\maketitle
\section{Introduction}
\label{sec:Introduction}
Quantum field theory in curved spacetime has interesting and partly counter-intuitive features such as the production of (pairs of) particles from the spacetime geometry itself \cite{Parker1969,Birrell1982,Mukhanov2007,Weinberg2008}, or interesting entanglement across spacetime horizons \cite{Gibbons1977,Page1993,Srednicki1993,Bombelli1986}. 
Other examples are Hawking radiation \cite{Hawking1975} and the related Unruh effect \cite{Unruh1981}.

Some of these features play a role in early time cosmology, e.g., during inflation, but cannot easily be investigated there in detail. It is therefore interesting to study other situations, such as in the context of condensed matter or atomic physics, which can be described by quantum field theory in curved spacetime. One may expect that an improved understanding that can be gained through the study of such analogous problems will also be profitable for the understanding of quantum fields in our Universe, or interesting phenomena of nonequilibrium quantum field theory in general.

In the present paper we shall be concerned with real, massless, relativistic scalar fields in $d=2+1$ spacetime dimensions. In particular we will study their quantization in spacetime geometries that are analogous to those found for $d=3+1$ in pioneering works by Friedmann \cite{Friedman1922,Friedman1924}, Lemaître \cite{Lemaitre1931}, Robertson \cite{Robertson1935,Robertson1936a,Robertson1936b} and Walker \cite{Walker1937} (FLRW). 

The class of FLRW universes in $d=2+1$ dimensions comprises three different possibilities for the spatial curvature $\kappa$, corresponding to a closed or spherical ($\kappa > 0$), a flat ($\kappa = 0$), and an open or hyperbolic ($\kappa < 0$) universe (see e.g. \cite{Ratra1995,Ratra2017} for studies of the $d=3+1$ dimensional FLRW universe with nonvanishing spatial curvature in the context of inflation). We develop here the means to analyze particle production in all three cases.

The reduced number of dimensions leads to some technical simplifications in the theoretical treatment, but also has advantages for experimental realizations. On the other side, a further reduction to a single spatial dimension would make spatial curvature trivial, and one would also encounter the problem that many choices for the metric in $d=1+1$ are related to the Minkowski metric by a conformal transformation, with the conformal group being of infinite dimension. This is different in $d=2+1$ or $d=3+1$ dimensions, where the conformal groups are only finite dimensional Lie groups. In this sense, $d=2+1$ dimensional spacetimes are the simplest nontrivial case where free, massless, relativistic quantum fields (constituting a conformal field theory) can be studied in detail.

Let us remark here that $d=2+1$ dimensional models have been extensively investigated in the context of the analog gravity program (see e.g. \cite{Barcelo2011,Visser2002,Novello2002} for comprehensive introductions). More precisely, it has been shown that the propagation of sound modes on top of the ground state of a Bose-Einstein condensate at zero temperature can be described by a relativistic Klein-Gordon equation with an underlying flat FLRW geometry \cite{Barcelo2003c,Fedichev2003,Fedichev2004,Fischer2004,Fischer2004b,Uhlmann2005,Calzetta2005,Weinfurtner2007,Weinfurtner2009,Prain2010}. 

In our companion paper \cite{BECPaper2022} we have generalized the aforementioned mapping to spatially curved FLRW universes, which can be realized by enclosing the Bose-Einstein condensate with suitable trapping profiles and varying the scattering length over time to simulate expanding, as well as contracting, FLRW cosmologies. In this way, observing cosmological particle production becomes accessible in table-top experiments. Exemplarily, we point to Refs.\ \cite{Eckel2018,Wittemer2019} for recent experimental approaches to simulate cosmological effects in the laboratory.

With the theoretical approach developed in the present paper we provide a basis for further theoretical, as well as experimental work on quantum field theory in curved spacetimes.

\paragraph*{The remainder of this paper is organized as follows.} In Sec.\ \ref{subsec:FLRWUniverses} we introduce the class and geometry of FLRW universes. We proceed then to the quantum field theory of a free scalar field (Sec.\ \ref{subsec:ScalarField}), leading to a Klein-Gordon equation in curved spacetime containing the Laplace-Beltrami operator in $d=2+1$ dimensions. The latter is discussed in Sec.\ \ref{subsec:LaplaceBeltrami}, allowing for canonical quantization of the quantum field, which is carried out in Sec.\ \ref{subsec:Quantization}. In Sec.\ \ref{sec:ParticleProduction}, we study cosmological particle production. In particular, we derive the spectrum of fluctuations and several two-point correlation functions (Sec.\ \ref{subsec:SpectrumAndCorrelators}). At last, we summarize our findings and provide an outlook in Sec.\ \ref{sec:Outlook}.

\paragraph*{Notation.} In this work, we employ natural units ${\hbar = c = k_\text{B} = 1}$ and use Minkowski space metric signature $(-,+,+)$. Furthermore, greek indices $\mu, \nu$ run from $0$ to $2$, while latin indices $i,j$ only run from $1$ to $2$.

\section{Cosmological spacetime with $d=2+1$ dimensions}
\label{subsec:FLRWUniverses}
\subsection{Geometry}
We first analyze geometric features of cosmological spacetimes in $d=2+1$ spacetime dimensions. The FLRW metric describing a homogeneous and isotropic universe can be formulated in comoving coordinates. These provide a natural foliation of spacetime, in spacelike hypersurfaces $\Sigma$ and a time coordinate $t$. The line element is
\begin{equation}
    \text{d} s^2 = - \text{d}t^2 + a^2(t) \gamma_{ij}\text{d}x^i \text{d}x^j,
    \label{eq:SyncLineElement}
\end{equation}
with a diagonal spatial metric $\gamma_{ij}$ on $\Sigma$ and a scale factor $a(t)$. 

The spatial line element is
\begin{equation}
    \gamma_{ij}\text{d}x^i \text{d}x^j = 
    \frac{\text{d} u^2}{1-\kappa u^2} + u^2\text{d}\varphi^2,
    \label{eq:SpatialLineElement1}
\end{equation}
where $\kappa$ parametrizes spatial curvature of a closed ($\kappa > 0$), flat ($\kappa = 0$), or open ($\kappa < 0$) universe, and $\varphi \in [0, 2 \pi)$ denotes an azimuthal angle. 

The nonvanishing Christoffel symbols are
\begin{equation}
\begin{split}
\Gamma^0_{11} &= \frac{a \dot a}{1-\kappa u^2}, \quad\quad 
\Gamma^0_{22} = a \dot a u^2, \\
\Gamma^1_{01} &=\Gamma^1_{10} = \Gamma^2_{02} = \Gamma^2_{20} = \frac{\dot a}{a}, \\
\Gamma^1_{11} &= \frac{\kappa u}{1-\kappa u^2}, \quad\quad 
\Gamma^1_{22} = - u (1-\kappa u^2), \\
\Gamma^2_{12} &= \Gamma^2_{21} = \frac{1}{u},
\end{split}
\end{equation}
with dots representing derivatives with respect to time, $t$.

The Ricci tensor is obtained by usual methods and has the nonvanishing components
\begin{equation}
\begin{split}
R_{00} & =  - 2 \frac{\ddot a}{a}, \\
R_{11} & = \frac{1}{1-\kappa u^2}[\ddot a a + \dot a^2+\kappa], \\
R_{22} & = u^2 [\ddot a a + \dot a^2+\kappa].
\end{split}
\end{equation}
The latter determines in $d=2+1$ dimensions also the Riemann curvature tensor. The Ricci scalar in turn is given by
\begin{equation}
R =  \frac{2\kappa+4\ddot a a+2\dot a^2}{a^2}.
\label{eq:RicciScalar}
\end{equation}

Curvature can also be parametrized as the extrinsic curvature tensor of a constant time surface $\Sigma$,
\begin{equation}
K_{ij} = \frac{1}{2} \partial_t g_{ij} = a \dot a \, \gamma_{ij},
\end{equation}
supplemented by the intrinsic curvature Ricci tensor
$$
R^{(2)}_{11} = \frac{\kappa}{1-\kappa u^2}, \quad\quad\quad R^{(2)}_{22} = u^2 \kappa,
$$
such that the intrinsic curvature scalar is twice the Gaussian curvature, $R^{(2)} = 2\kappa / a^2$.

Another interesting object in $d=2+1$ dimensions is the Cotton tensor $C_{\mu\nu\rho}$, which vanishes precisely when the metric is conformally flat. For the metric specified by \eqref{eq:SyncLineElement} and \eqref{eq:SpatialLineElement1} it vanishes, $C_{\mu\nu\rho}=0$, so that the entire class of spacetime geometries is actually conformally flat.

The Einstein tensor $G_{\mu\nu} = R_{\mu\nu} - (R/2) g_{\mu\nu}$ has the nonvanishing components
\begin{equation}
G_{00} = \frac{\kappa + \dot a^2}{a^2}, \quad\quad G_{ij} = - a \ddot a \gamma_{ij},
\end{equation}
while the energy-momentum tensor for a fluid in a homogeneous and isotropic state is
\begin{equation}
T^{\mu\nu} = \epsilon u^\mu u^\nu + (p+\pi_\text{bulk}) [u^\mu u^\nu + g^{\mu\nu}],
\end{equation}
where $u^\mu=(1,0,0)$ is the fluid velocity, $\epsilon$ the energy density, and $p+\pi_\text{bulk}$ the effective pressure. The analog of Einsteins equations in $d=2+1$ dimensions,
\begin{equation}
G_{\mu\nu} =  8\pi G_\text{N} T_{\mu\nu},
\end{equation}
would imply the two Friedmann equations
\begin{equation}
\frac{\kappa + \dot a^2}{a^2} = 8\pi G_\text{N} \epsilon, \quad\quad -\frac{a \ddot a}{a^2} = 8\pi G_\text{N} (p+\pi_\text{bulk}). 
\label{eq:Friedmann}
\end{equation}
In this context one should note that $G_\text{N}$ is not the real Newton coupling but rather its analog in $d=2+1$ dimensions. It differs from the former in several ways, for example it has a different engineering dimension.

It is useful to also note the conservation law 
\begin{equation}
\dot \epsilon + 2 \frac{\dot a}{a} (\epsilon + p + \pi_\text{bulk}) = 0.
\end{equation}
For a matter dominated universe where $p+\pi_\text{bulk}=0$, this leads to $\epsilon(t) = \epsilon_0 a_0^2 / a(t)^2$. Similarly, for a radiation dominated situation with $p+\pi_\text{bulk}=\epsilon / 2$ one finds $\epsilon(t) = \epsilon_0 a_0^3 / a(t)^3$, and finally, for a situation dominated by a cosmological constant with $p+\pi_\text{bulk}=-\epsilon$ one would have constant energy density, $\epsilon(t) = \epsilon_0$. Using \eqref{eq:Friedmann} one finds then for the matter dominated universe $a(t) = a_0 t$ such that $\dot a$ is constant. Assuming $\kappa=0$, the radiation dominated universe has $a(t) = a_0 t^{2/3}$, and the one dominated by a cosmological constant, as usual, $a(t)= a_0 e^{H t}$. We note that nonvanishing spatial curvature modifies these latter relations somewhat.

Finally, a universe without any matter content, ${\epsilon = p+\pi_\text{bulk}=0}$, would only fulfill Einsteins equations for $\kappa \leq 0$ and then have a linear scale factor of the form $
a(t) = \sqrt{-\kappa}\, t$.

\subsection{Symmetries and Killing vector fields}

Symmetries of a curved spacetime geometry are described by the {\it conformal Killing equation}
\begin{equation}
\begin{split}
& g_{\lambda \sigma}(x) \partial_\rho \xi^\lambda(x) + g_{\rho \lambda}(x) \partial_\sigma \xi^\lambda(x) \\
& + \xi^\lambda(x) \partial_\lambda g_{\rho \sigma}(x) + \omega(x) g_{\rho \sigma}(x) = 0.
\end{split}
\label{eq:conformal_killing_eq}
\end{equation}
A solution $\xi^\mu(x)$ is called conformal Killing vector field. It parametrizes an infinitesimal change of coordinates, $x^\mu \to x^{\prime\mu} = x^\mu + \epsilon \xi^\mu(x)$, for which the metric is invariant up to an overall conformal factor $\Omega^2(x) = 1+\epsilon \omega(x)$. A solution $\xi^\mu(x)$ with $\omega(x)=0$ is called a Killing vector field and it parametrizes a change of coordinates for which the metric does not change at all. This is obviously a stronger condition.

Let us now take the metric to be of the form \eqref{eq:SyncLineElement} with spatial line element as in \eqref{eq:SpatialLineElement1}. The radial coordinate $u$ may be further specified depending on the spatial curvature $\kappa$. That is, we use an angle $\theta \in [0,\pi)$ for a closed universe, a pseudoangle $\sigma\in[0,\infty)$ for an open universe, and a radial coordinate $u\in[0,\infty)$ for a flat universe, so that the spatial line element becomes
\begin{equation}
    \gamma_{ij}\text{d}x^i \text{d}x^j = \begin{cases}
    \frac{1}{|\kappa|}  (\text{d}\theta^2 + \sin^2\theta \, \text{d}\varphi^2) & \text{for}\hspace{0.3cm} \kappa > 0, \\
    \text{d} u^2 + u^2\text{d}\varphi^2  & \text{for}\hspace{0.3cm} \kappa = 0,\\
    \frac{1}{|\kappa|}  ( \text{d}\sigma^2 + \sinh^2 \sigma \, \text{d}\varphi^2) & \text{for}\hspace{0.3cm} \kappa < 0.
    \end{cases}
    \label{eq:SpatialLineElement}
\end{equation}

In the case $\kappa>0$ there are three spatial Killing vector fields corresponding to the three infinitesimal rotations of the sphere. They can be taken to be
\begin{equation}
\begin{pmatrix} 0 \\ \sin\varphi \\ \cot\theta \cos\varphi \end{pmatrix}, \quad
\begin{pmatrix} 0 \\ \cos\varphi \\ -\cot\theta \sin\varphi \end{pmatrix}, \quad
\begin{pmatrix} 0 \\ 0 \\ 1 \end{pmatrix}.
\end{equation}
Together with the commutator of vector fields, they generate the Lie algebra of SO$(3)$ or SU$(2)$. Similarly, for $\kappa<0$ one has again three spatial Killing vector fields 
\begin{equation}
\begin{pmatrix} 0 \\ \sin\varphi \\ \coth\sigma \cos\varphi \end{pmatrix}, \quad
\begin{pmatrix} 0 \\ \cos\varphi \\ -\coth\sigma \sin\varphi \end{pmatrix}, \quad
\begin{pmatrix} 0 \\ 0 \\ 1 \end{pmatrix},
\end{equation}
which generate the Lie algebra of SO$(2,1)$ or SU$(1,1)$. Finally, in case of $\kappa=0$ one has the tree spatial Killing vector fields
\begin{equation}
\begin{pmatrix} 0 \\ \sin\varphi \\ \cos\varphi \end{pmatrix}, \quad
\begin{pmatrix} 0 \\ \cos\varphi \\ -\sin\varphi \end{pmatrix}, \quad
\begin{pmatrix} 0 \\ 0 \\ 1 \end{pmatrix},
\end{equation}
corresponding to two translations and a rotation in the Euclidean plane with symmetry group E$(2)$.

One may also check that the metric specified by \eqref{eq:SyncLineElement} has no timelike Killing vector field, except for constant scale factor, $\dot a=0$. However, it has a timelike conformal Killing vector field $\xi^\mu=(a,0,0)$ with $\omega=-2\dot a$.

\section{Scalar quantum field in an FLRW universe}
\label{sec:ScalarFieldInFLRWUniverse}
We now discuss the quantization procedure for a free  massless scalar field $\phi$ in a $d=2+1$ dimensional (spatially curved) FLRW universe step by step.

\subsection{Free relativistic scalar field}
\label{subsec:ScalarField}
The action for a real massless scalar field in a $d=2+1$ dimensional geometry can be written as
\begin{equation}
    \Gamma [\phi] = -\frac{1}{2} \int \text{d} t \, \text{d}^2 x \, \sqrt{g} \, g^{\mu\nu} \partial_\mu \phi \partial_\nu \phi,
    \label{eq:QEACurvedSpacetime}
\end{equation}
where $ \sqrt{g} \equiv \sqrt{- \det (g_{\mu \nu})}$ denotes the determinant of the metric. The metric determinant becomes with \eqref{eq:SpatialLineElement}
\begin{equation}
    \begin{split}
        \sqrt{g} = a^2(t)\times \begin{cases}
        \frac{\sin\theta}{\abs{\kappa}} &\text{for}\hspace{0.2cm} \kappa > 0, \\
        u &\text{for}\hspace{0.2cm} \kappa = 0, \\
        \frac{\sinh\sigma}{\abs{\kappa}} &\text{for}\hspace{0.2cm} \kappa < 0.
        \end{cases}
    \end{split}
    \label{eq:Determinant}
\end{equation}

The equations of motion for the field $\phi$ can be obtained from a variation of the action \eqref{eq:QEACurvedSpacetime} with respect to $\phi$ using Eq.\ \eqref{eq:SpatialLineElement}, which leads to a Klein-Gordon equation in curved spacetime
\begin{equation}
    \begin{split}
        0 &= \partial_\mu \left(\sqrt{g} \, g^{\mu \nu} \, \partial_\nu \phi \right) \\
        &= 2 a(t)\dot{a}(t)\dot{\phi} + a^2(t) \ddot{\phi} - \Delta \phi.
    \end{split}
    \label{eq:KleinGordonFLRW}
\end{equation}
Here, $\Delta$ denotes the Laplace-Beltrami operator in the spatial geometry fixed by Eq.\ \eqref{eq:SpatialLineElement1}. Its form depends on the spatial curvature $\kappa$, as is discussed in detail in the next subsection.

\subsection{Laplace-Beltrami operator in $\mathbf{d=2+1}$ dimensions}
\label{subsec:LaplaceBeltrami}
In general, we wish to solve Eq.\ \eqref{eq:KleinGordonFLRW} for all three classes of spatial curvature $\kappa$, with the choice of coordinates according to Eq.\ \eqref{eq:SpatialLineElement}. To that end, we consider the corresponding Laplace-Beltrami operator adapted to the spatial geometry defined by $\kappa$, which reads \cite{Ratra1995,Ratra2017,Argyres1989}
\begin{equation}
    \Delta = \begin{cases}
    \abs{\kappa} \Big[ \frac{
    1}{\sin \theta}\partial_{\theta}\left(\sin \theta \, \partial_{\theta} \right) + \frac{
    1}{\sin^2\theta}\partial_{\varphi}^2 \Big] &\text{for}\hspace{0.2cm} \kappa > 0, \\
    \partial^2_u + \frac{1}{u} \partial_u + \frac{1}{u^2} \partial_{\varphi}^2 &\text{for}\hspace{0.2cm} \kappa = 0, \\
    \abs{\kappa} \Big[\frac{
    1}{\sinh \sigma}\partial_{\sigma}\left(\sinh \sigma \, \partial_{\sigma} \right) + \frac{
    1}{\sinh^2\sigma}\partial_{\varphi}^2 \Big] &\text{for}\hspace{0.2cm} \kappa < 0.
    \end{cases}
    \label{eq:LaplaceOperator}
\end{equation}

In the following it will be convenient to introduce momentum modes that diagonalize the Laplace-Beltrami operator $\Delta$. We use a radial wave number $k$ or $l$, related by $k=\sqrt{\abs{\kappa}}l$, and an azimuthal wave number $m$, with the ranges
\begin{equation}
    \begin{split}
        &l \in \mathbb{N}_0, m \in \{-l,...,l \} \hspace{0.3cm} \text{for} \hspace{0.2cm} \kappa > 0, \\
        &k \in \mathbb{R}^+_0, m \in \mathbb{Z} \hspace{1.36cm} \text{for} \hspace{0.2cm} \kappa = 0, \\
        &l \in \mathbb{R}^+_0, m \in \mathbb{Z} \hspace{1.44cm} \text{for} \hspace{0.2cm} \kappa < 0.
    \end{split}
    \label{eq:MomentumVector}
\end{equation}
In momentum space, the Laplace-Beltrami operator becomes diagonal, in the sense that it fulfills an eigenvalue equation
\begin{equation}
    \Delta \mathcal{H}_{km} (u,\varphi) = h (k) \, \mathcal{H}_{km} (u,\varphi),
    \label{eq:LaplaceEigenvalueEquation}
\end{equation}
with corresponding sets of complete and orthonormal eigenfunctions
\begin{equation}
    \mathcal{H}_{km} (u,\varphi) = \begin{cases}
    Y_{lm}(\theta,\varphi) &\text{for} \hspace{0.2cm} \kappa > 0, \\
    X_{km} (u, \varphi) &\text{for} \hspace{0.2cm} \kappa = 0, \\
    W_{lm} (\sigma, \varphi) &\text{for} \hspace{0.2cm} \kappa < 0,
    \end{cases}
    \label{eq:LaplaceEigenfunctions}
\end{equation}
and eigenvalues $h(k)$,
\begin{equation}
    h(k) = \begin{cases}
    - k (k+\sqrt{\abs{\kappa}}) & \text{for}\hspace{0.3cm} \kappa > 0, \\
    - k^2 & \text{for}\hspace{0.3cm} \kappa = 0, \\
    - \left( k^2 + \frac{1}{4}\abs{\kappa}\right) & \text{for}\hspace{0.3cm} \kappa < 0.
    \end{cases}
    \label{eq:LaplaceEigenvalues}
\end{equation}
Note that $h(k)$ vanishes for a vanishing radial wave number $k=0$, unless the spatial curvature is negative, i.e. $\kappa < 0$.

One should remark here that the eigenvalue spectrum of the Laplace-Beltrami operator depends in general on boundary conditions. For $\kappa \leq 0$, we are dealing with an infinite space where boundary conditions are set asymptotically, while for $\kappa >0$ the space is finite and we have chosen it to have the topology of a sphere. It is expected that other boundary conditions would modify Eq.\ \eqref{eq:LaplaceEigenvalues}.

We turn now to an analysis of the sets of eigenfunctions: concretely, $Y_{lm}(\theta,\varphi)$ are a modified version of the spherical harmonics,
\begin{equation}
    Y_{lm}(\theta,\varphi) = \sqrt{\frac{(l-m)!}{(l+m)!}} \, e^{im\varphi} \, P_{lm}(\cos \theta),
    \label{eq:SphericalHarmonics}
\end{equation}
with $P_{lm}(\cos\theta) = (-1)^mP^m_l(\cos\theta)$ being the associated Legendre polynomials. In contrast to the standard definition of the spherical harmonics, we have introduced the sign $(-1)^m$ and omitted a factor $\sqrt{4\pi}/\sqrt{2l+1}$. 

Moreover, $X_{k m} (u, \varphi)$ are related to Bessel functions of the first kind via
\begin{equation}
    X_{km} (u, \varphi) = e^{i m \varphi} \, J_m (k u),
    \label{eq:EigenfunctionFlat}
\end{equation}
while the eigenfunctions $W_{lm} (\sigma, \varphi)$ are given by
\begin{equation}
    \begin{split}
        W_{lm} (\sigma, \varphi) = &(-i)^m \frac{\Gamma(il+1/2)}{\Gamma(il+m+1/2)} \\
        &\times e^{im\varphi} P^m_{il-1/2}\left(\cosh \sigma\right),    
    \label{eq:EigenfunctionOpen}
    \end{split}
\end{equation}
wherein $P^m_{il-1/2}\left(\cosh \sigma\right)$ are conical functions corresponding to analytically continued Legendre functions. 

The functions specified in Eqs.\ \eqref{eq:SphericalHarmonics}, \eqref{eq:EigenfunctionFlat}, and \eqref{eq:EigenfunctionOpen} are normalized with respect to a scalar product, which will be discussed next. For the closed FLRW universe, the adapted version of the spherical harmonics $Y_{lm}(\theta,\varphi)$ defined in Eq.\ \eqref{eq:SphericalHarmonics} form an orthonormal set,
\begin{equation}
    \begin{split}
        &\left(Y_{lm},Y_{l^{\prime}m^{\prime}}\right)\\ &= \frac{1}{\abs{\kappa}} \int_0^{\pi}\text{d}\theta \sin\theta \int_0^{2\pi} \text{d}\varphi \, Y^{*}_{lm}(\theta,\varphi) Y_{l^{\prime}m^{\prime}}(\theta,\varphi)\\
        &= \frac{1}{\abs{\kappa}} \frac{2\pi}{l+1/2} \delta_{mm^{\prime}}\delta_{ll^{\prime}},
    \end{split}
\end{equation}
and fulfill the completeness relation
\begin{equation}
    \begin{split}
        &\sum_{l=0}^{\infty} \abs{\kappa}\frac{l+1/2}{2\pi} \sum_{m=-l}^{l} Y_{lm}(\theta,\varphi) Y_{lm}^*(\theta^{\prime},\varphi^{\prime})\\
        &= \frac{\abs{\kappa}}{\sin\theta} \delta(\theta-\theta^{\prime}) \delta(\varphi-\varphi^{\prime}).
    \end{split}
\end{equation}
The eigenfunctions for a flat FLRW universe $X_{km}(u,\varphi)$ as defined in Eq.\ \eqref{eq:EigenfunctionFlat} also form an orthonormal set
\begin{equation}
\begin{split}
    &\left(X_{km},X_{k^{\prime}m^{\prime}}\right) \\
    &= \int_0^{\infty}\text{d}u \,u \int_0^{2\pi} \text{d}\varphi \, X^{*}_{km}(u,\varphi) X_{k^{\prime}m^{\prime}}(u,\varphi)\\
    &= \frac{2\pi}{k}\delta_{mm^{\prime}}\delta(k-k^{\prime}),
\end{split}
\end{equation}
while their completeness relation reads
\begin{equation}
\begin{split}
    &\int_0^{\infty} \frac{\text{d}k}{2\pi} k \sum_{m=-\infty}^{\infty} X_{km}(u,\varphi) X_{km}^*(u^{\prime},\varphi^{\prime})\\
    &= \frac{1}{u}\delta(u-u^{\prime}) \delta(\varphi-\varphi^{\prime}).
\end{split}
\end{equation}
Finally, the orthonormality relation for the eigenfunctions of an open FLRW universe $W_{lm}(\sigma,\varphi)$, as defined in Eq.~\eqref{eq:EigenfunctionOpen}, is given by
\begin{equation}
    \begin{split}
        &\hspace{-0.15cm}\big(W_{lm}, W_{l^{\prime}m^{\prime}}\big) \\ &\hspace{-0.15cm}=\frac{1}{\abs{\kappa}}\int_0^{\infty}\text{d}\sigma \sinh\sigma \int_0^{2\pi} \text{d}\varphi \, W^{*}_{lm}(\sigma,\varphi) W_{l^{\prime}m^{\prime}}(\sigma,\varphi) \\
        &\hspace{-0.15cm}= \frac{1}{\abs{\kappa}} \frac{2\pi}{l\tanh(\pi l)} \delta_{mm^{\prime}}\delta(l-l^{\prime}),
    \end{split}
\end{equation}
and the completeness relation is here
\begin{equation}
    \begin{split}
        &\int_0^{\infty} \frac{\text{d}l}{2\pi} \abs{\kappa} l \tanh(\pi l) \sum_{m=-\infty}^{\infty} W_{lm}(\sigma,\varphi)W_{lm}^*(\sigma^{\prime},\varphi^{\prime})\\
        &= \frac{\abs{\kappa}}{\sinh \sigma} \delta(\sigma-\sigma^{\prime}) \delta(\varphi-\varphi^{\prime}).
    \end{split}
\end{equation}
In the following we will use these mode decompositions for field quantization.

\subsection{Field quantization and mode functions}
\label{subsec:Quantization}
The bosonic field $\phi$ can be quantized by promoting it to an operator, such that it fulfills the (equal time) commutation relations,
\begin{equation}
    \begin{split}
        &[\phi(t, u,\varphi), \pi(t,u^{\prime},\varphi^{\prime})] \\
        &= i \delta(\varphi-\varphi^{\prime})\times
        \begin{cases}
        \delta(\theta-\theta') & \text{for}\hspace{0.3cm} \kappa > 0,\\
        \delta(u-u')& \text{for}\hspace{0.3cm} \kappa = 0,\\
        \delta(\sigma-\sigma')& \text{for}\hspace{0.3cm} \kappa < 0,
        \end{cases}
    \end{split}
    \label{eq:FLRWCommutationRelationsFields}
\end{equation}
with
\begin{equation}
    \begin{split}
        \pi(t,u,\varphi) = \frac{\delta \Gamma_2 [\phi]}{\delta \dot{\phi}} &= \sqrt{g} \dot{\phi},
    \end{split}
\end{equation}
being the conjugate momentum field. 

The quantum field operator $\phi$ can be expanded using the eigenfunctions of the Laplace-Beltrami operator \eqref{eq:LaplaceEigenfunctions} for the three different universes, which leads to
\begin{equation}
    \begin{split}
        &\phi(t, u,\varphi)\\ 
        &= \int_{k,m}\left[\hat{a}_{km} \mathcal{H}_{km} (u,\varphi) v_k (t) + \hat{a}^{\dagger}_{km} \mathcal{H}_{km}^{*} (u,\varphi) v_k^* (t) \right],
    \end{split}
    \label{eq:ModeExpansion}
\end{equation}
(analogously for the conjugate momentum field $\pi(t, u,\varphi)$), where we introduced momentum integrals via
\begin{equation}
    \int_{k,m}  = \begin{cases}
    \sum_{l=0}^{\infty} \abs{\kappa}\frac{l+1/2}{2\pi} \sum_{m=-l}^{l} &\text{for} \hspace{0.2cm} \kappa > 0, \\
    \int \frac{\text{d}k}{2\pi} \, k \sum_{m=-\infty}^{\infty} &\text{for} \hspace{0.2cm} \kappa = 0,\\
    \int \frac{\text{d}l}{2\pi} \,\abs{\kappa} l \tanh(\pi l) \sum_{m=-\infty}^{\infty} &\text{for} \hspace{0.2cm} \kappa < 0,
    \end{cases}
\end{equation}
and creation and annihilation operators $\hat{a}^{\dagger}_{km}$ and $\hat{a}_{km}$, respectively, which obey the usual bosonic commutation relations
\begin{equation}
    \begin{split}
        &[\hat{a}_{km}^{\dagger},\hat{a}_{k^{\prime}m^{\prime}}^{\dagger}] = [\hat{a}_{km},\hat{a}_{k^{\prime}m^{\prime}}] = 0, \\
        &[\hat{a}_{km},\hat{a}_{k^{\prime}m^{\prime}}^{\dagger}] = 2\pi \delta_{mm^{\prime}} \begin{cases}
         \frac{\delta_{ll^{\prime}}}{\abs{\kappa}(l+1/2)} & \text{for}\hspace{0.3cm} \kappa > 0, \\
            \frac{\delta(k-k^{\prime})}{k} & \text{for}\hspace{0.3cm} \kappa = 0,  \\
            \frac{\delta(l-l^{\prime})}{\abs{\kappa}l \tanh(\pi l)} & \text{for}\hspace{0.3cm} \kappa < 0.
        \end{cases}
        \label{eq:CommutationRelationsBosonicOperators}
    \end{split}
\end{equation}
We also introduced time-dependent mode functions $v_{k} (t)$ and $v_k^{*} (t)$.
Using the mode expansion \eqref{eq:ModeExpansion} in the Klein-Gordon equation \eqref{eq:KleinGordonFLRW} yields the so-called mode equation
\begin{equation}
    \ddot{v}_k (t) + 2\frac{\dot{a}(t)}{a(t)}\dot{v}_k(t) - \frac{h(k)}{a^2(t)} v_k(t) = 0,
    \label{eq:ModeEquation}
\end{equation}
where the dependence on the spatial curvature $\kappa$ is entirely included in the eigenvalues of the Laplace-Beltrami operator $h(k)$. 

Furthermore, the Wronskian of the mode functions $v_k$ and $v_k^*$ fulfills a normalization condition
\begin{equation} 
     \text{Wr}[v_k, v_k^*] = a^2(t) \left[  v_k \dot{v}_{k}^{*} - \dot{v}_{k} v_k^{*}  \right] = i,
     \label{eq:ModeRelation}
\end{equation}
which is a consequence of the canonical commutation relations \eqref{eq:FLRWCommutationRelationsFields} and \eqref{eq:CommutationRelationsBosonicOperators} and the orthonormality properties of the functions $\mathcal{H}_{km} (u,\varphi)$. One may check that the normalization condition \eqref{eq:ModeRelation} is fulfilled at all times provided that it is fulfilled at one instance of time.

\section{Particle Production}
\label{sec:ParticleProduction}
We now introduce Bogoliubov transformations to describe particle production within the FLRW cosmology paradigm for all three types of spatial curvature. In this context, we discuss the power spectrum of fluctuations and several variants of two-point correlation functions.

For a time-dependent scale factor $a(t)$, or more generally in the absence of a timelike Killing vector field, the notions of particles and quantum field theoretic vacuum are not unique \cite{Parker1969,Birrell1982}. An interesting situation arises also when the initial and final state have stationary scale factor, but $a(t)$ varies at intermediate time. This leads to particle production and plays a role for example in early time cosmology, such as an inflationary epoch where $a(t)$ evolved strongly. We now discuss this phenomenon in more detail for the $d=2+1$ dimensional spacetimes introduced in Sec. \ref{subsec:FLRWUniverses}.

\subsection{Bogoliubov transformations}
\label{subsec:BogoliubovCoefficients}
\begin{figure}[t!]
    \includegraphics[width=0.47\textwidth]{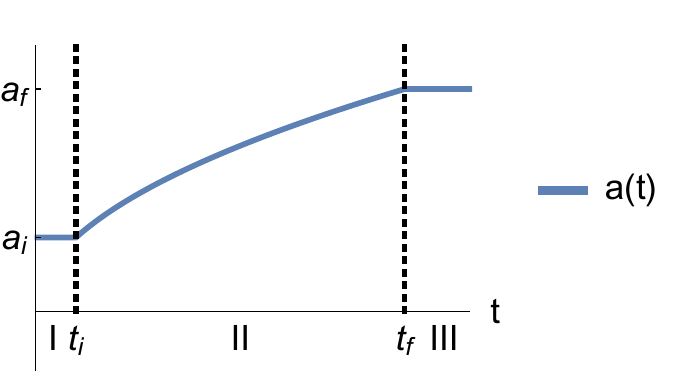}
    \caption{The time evolution of the scale factor $a(t)$ (blue solid line) is shown for the three regions I, II and III. For $t < t_\text{i}$, the scale factor is constant $a(t<t_{\text{i}} = a_{\text{i}})$, while it becomes time-dependent in region II. Exemplary, we show a polynomial expansion with $\gamma = 1/2$ (the two constants were set to $Q = 1$ and $t_0=0$). In region III, the scale factor is again stationary, i.e. $a (t > t_\text{f}) = a_{\text{f}}$.} 
    \label{fig:Expansion}
\end{figure}

We may consider a scenario (cf. Fig.\ \ref{fig:Expansion}) where the scale factor is constant, $a(t)=a_\text{i}$, in some time interval (which we refer to as region I) up to $t_{\text{i}}$. Then, the mode functions $v_k$ in region I are given by oscillatory modes
\begin{equation}
    v_k^{\text{I}} (t) = \frac{\exp (-i \omega_k^{\text{I}} t)}{a_\text{i} \sqrt{2 \omega_k^{\text{I}}}},
    \label{eq:planewavetime}
\end{equation}
with a positive frequency $\omega_k^{\text{I}} = \sqrt{-h(k)/a^2_\text{i}}$, while the creation and annihilation operators $\hat a_{km}^\dagger$ and $\hat a_{km}$, respectively, create and annihilate quantum excitations. This allows us to unambiguously define a vacuum state $\ket{\Omega}$ via
\begin{equation}
    \hat a_{km}\ket{\Omega} = 0.
    \label{eq:defOmegaState}
\end{equation}
In a second time interval $t_\text{i} < t < t_\text{f}$ (or region II), the scale factor $a(t)$ attains a time dependence, while it becomes constant again for $t \ge t_{\text{f}}$ (region III in the following). In the time-dependent region II, we focus on polynomial scale factors, such that for all times the scale factor may be written as
\begin{equation}
    a(t) = \begin{cases}
            a_\text{i}  & \text{for } t \leq t_\text{i}, \\
            Q |t -t_0|^\gamma & \text{for } t_\text{i} < t < t_\text{f}, \\
            a_\text{f}  & \text{for } t_\text{f} \leq t,
            \end{cases}
    \label{eq:ScaleFactorPolynomial}
\end{equation}
where $Q$ and $t_0$ are free parameters, and $\gamma$ specifies the polynomial degree of the expansion for times between $t_\text{i}$ and $t_{\text{f}}$.

In region II, the mode functions can be obtained from the mode equation \eqref{eq:ModeEquation} together with two initial conditions at $t=t_\text{i}$ stemming from requiring continuity in the solution \eqref{eq:planewavetime}. 

Finally, in region III corresponding to $t \ge t_\text{f}$, the solutions to the mode equation \eqref{eq:ModeEquation} are again of oscillatory form,
\begin{equation}
    u_k^{\text{III}} (t) = \frac{\exp(-i \omega_k^{\text{III}} t)}{a_\text{f} \sqrt{2 \omega_k^{\text{III}}}},
\end{equation} 
with positive frequencies $\omega_k^{\text{III}} = \sqrt{-h(k)/a_\text{f}^2}$ being now related to the final scale factor $a(t \ge t_{\text{f}}) = a_{\text{f}}$. Quasiparticle excitations are related to a second set of creation and annihilation operators denoted by $\hat b^\dagger_{km}$ and $\hat b_{km}$, respectively, with a corresponding ``vacuum'' state $\ket{\Psi}$ fulfilling
\begin{equation}
    \hat b_{km} \ket{\Psi} = 0.
\end{equation}

At this point, a crucial observation is that the solution to the mode equation $v_k (t)$ for all three regions can in region III be written as a linear superposition of positive and negative frequency solutions, i.e.
\begin{equation}
    u_k =\alpha_k v_k + \beta_{k} v_k^{*}, \quad\quad\quad v_k = \alpha^*_k u_k - \beta_k u^*_k, 
    \label{eq:BogoliubovTransformationModes}    
\end{equation}
where $\alpha_k$ and $\beta_k$ denote the so-called Bogoliubov coefficients, which are time-independent and in general complex-valued. As a consequence of both sets of mode functions $v_k (t)$ and $u_k (t)$ being normalized according to \eqref{eq:ModeRelation}, the Bogoliubov coefficients fulfill the well-known relation 
\begin{equation}
    \abs{\alpha_k}^2 - \abs{\beta_k}^2 = 1.    
\end{equation}
The latter can be rewritten using the Wronskian defined in \eqref{eq:ModeRelation}, leading to
\begin{equation}
    \alpha_k = \text{Wr}[u_k, v_k^*]/i, \quad\quad\quad \beta_k = \text{Wr}[u_k, v_k]/i.
    \label{eq:BogoliubovInTermsOfWroskian}
\end{equation}
As with the mode functions $v_k (t)$ and $u_k (t)$, one can relate the two sets of creation and annihilation operators,
\begin{equation}
    \hat{b}_{km} =  \alpha_k^* \hat{a}_{km} - \beta_k^* (-1)^m \hat{a}^{\dagger}_{k,-m}.
    \label{eq:BogoliubovTransformationOperators}
\end{equation}
Hence, the initial vacuum state $\ket{\Omega}$ carries a finite particle content with respect to the excitations created by the operator $b_{k m}^{\dagger}$ this comes about given the time-dependence of the scale factor $a(t)$ in region II, and is the phenomenon referred to as \textit{particle production}.

To analyze this phenomenon one solves the mode equation \eqref{eq:ModeEquation} for the three regions I-III, and thereafter identifies the Bogoliubov coefficients $\alpha_k$ and $\beta_k$ through Eq.\ \eqref{eq:BogoliubovTransformationModes} (or equivalently Eq.\ \eqref{eq:BogoliubovInTermsOfWroskian}). As we will see in forthcoming sections, the knowledge of $\alpha_k$ and $\beta_k$ suffices to compute all relevant quantities related to particle production stemming from a time dependent scale factor.

\subsection{Time evolution inside and outside the horizon}
\label{subsec:Horizons}

In this subsection we analyze the implications of the mode equation \eqref{eq:ModeEquation} in more detail, and distinguish in particular between large wave vectors corresponding to small scales inside the Hubble horizon, and small wave numbers or large length scales outside of the Hubble horizon. For this discussion it is convenient to introduce the conformal time $\eta$ through
\begin{equation}
    \text{d}t=a(t) \text{d}\eta,
\end{equation}
and to introduce the rescaled mode function
\begin{equation}
    w_k(\eta) = \sqrt{a(t)} v_k(t).
\end{equation}
Equation \eqref{eq:ModeEquation} can now be written as
\begin{equation}
\frac{\text{d}^2}{\text{d}\eta^2} w_k(\eta) + \left[-h(k) -\left(\frac{1}{4} -\frac{1}{2}q\right)\mathcal{H}^2\right] w_k(\eta) = 0.
\label{eq:ModifiedModeEquation}
\end{equation}
We introduce here the conformal Hubble rate
\begin{equation}
\mathcal{H} = aH =\dot{a},
\end{equation}
and the deceleration parameter
\begin{equation}
q=-\frac{\ddot{a}a}{\dot{a}^2}.
\end{equation}

Interestingly for $\kappa \leq 0$ one can write, using \eqref{eq:LaplaceEigenvalues} and \eqref{eq:RicciScalar}, the mode equation \eqref{eq:ModifiedModeEquation} as
\begin{equation}
\begin{split}
\frac{\text{d}^2}{\text{d}\eta^2} w_k(\eta) + \left[k^2 - \frac{\kappa}{4} - \frac{\ddot a a}{2} - \frac{\dot a^2}{4}  \right] w_k(\eta) & = \\
\frac{\text{d}^2}{\text{d}\eta^2} w_k(\eta) + \left[k^2 -\frac{a^2}{8} R\right] w_k(\eta) & = 0,
\end{split}
\label{eq:ModifiedModeEquation2}
\end{equation}
which shows directly how spacetime curvature enters this equation in the form of the Ricci scalar $R$.

In the following we consider specifically the power-law scale factors in Eq.\ \eqref{eq:ScaleFactorPolynomial} with $t_0=0$ and $t>0$, where one has
\begin{equation}
    \mathcal{H} = \gamma Q t^{\gamma-1},
\end{equation}
and
\begin{equation}
    q=\frac{1}{\gamma}-1.
\end{equation}
For $\gamma>1$ the conformal Hubble rate $\mathcal{H}$ is increasing with time, and $q<0$. This corresponds to an accelerating expansion. In contrast, when $\gamma<1$ the conformal Hubble rate $\mathcal{H}$ is decreasing and $q>0$, corresponding to a decelerated expansion. In the marginal case of $\gamma=1$ one has the so-called coasting universe with constant conformal Hubble rate.

Another interesting transition happens at $\gamma=2/3$ and $q=1/2$. For $\gamma>2/3$ and $q<1/2$ the round bracket in \eqref{eq:ModifiedModeEquation} is positive, so that $(1/4-q/2)\mathcal{H}^2$ is like a negative ``effective mass squared''. In contrast, for $\gamma<2/3$ and $q>1/2$ this ``effective mass squared'' is positive, and it vanishes at the transition point with $\gamma=2/3$ and $q=1/2$, as well as for a stationary universe with $\mathcal{H}=0$.

\begin{figure}
\centering
\includegraphics[height=0.39\textwidth]{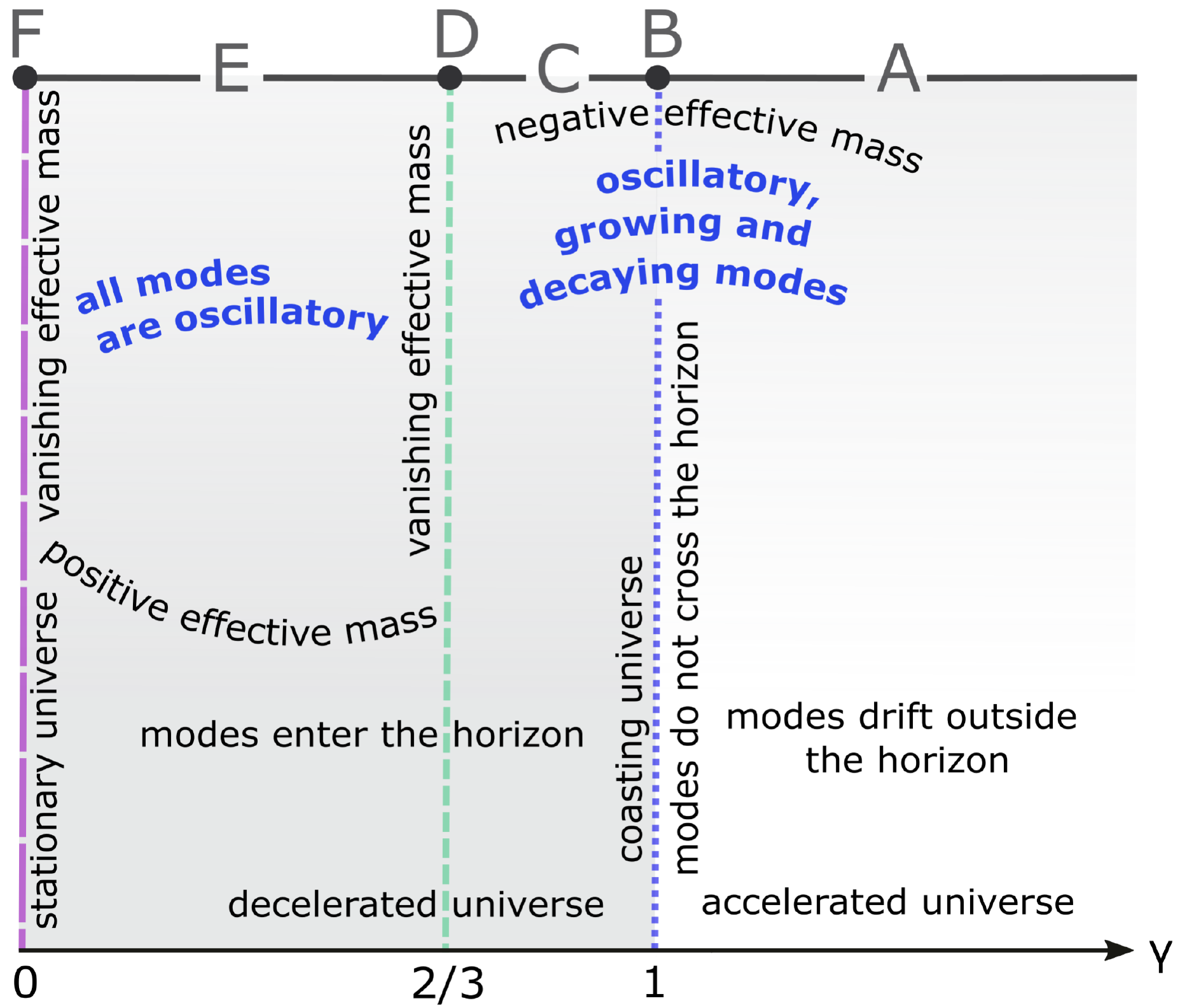}
    \caption{Different regimes for evolving mode functions in a $d=2+1$ dimensional universe with power-law scale factor $a(t) \sim t^{\gamma}$. Indicated in capital letters are the different scenarios for power-law expansions specified in the text.}
    \label{fig:HorizonModes}
\end{figure}

In summary, one has the following scenarios for power-law expansions:
\begin{itemize}
\item[(A)] Accelerating universe with $q<0$, $\gamma>1$. Negative Hubble-curvature ``effective mass squared'' with increasing magnitude due to increasing conformal Hubble rate $\mathcal{H}$.
\item[(B)] Coasting universe with $q=0$, $\gamma=1$. Negative Hubble-curvature ``effective mass squared'' with constant magnitude due to constant conformal Hubble rate $\mathcal{H}$.
\item[(C)] Decelerating universe with $0<q<1/2$, $2/3<\gamma<1$. Negative Hubble-curvature ``effective mass squared'' with decreasing magnitude due to decreasing conformal Hubble rate $\mathcal{H}$.
\item[(D)] Decelerating universe with $q=1/2$, $\gamma=2/3$. Vanishing Hubble-curvature ``effective mass squared'' at decreasing conformal Hubble rate $\mathcal{H}$.
\item[(E)] Decelerating universe with $q>1/2$, $\gamma<2/3$. Positive Hubble-curvature ``effective mass squared'' at decreasing conformal Hubble rate $\mathcal{H}$.
\item[(F)] Stationary universe with $\gamma=0$ and vanishing conformal Hubble rate $\mathcal{H}$.
\end{itemize}
An overview over these cases is also given in Fig.\ \ref{fig:HorizonModes}. A similar classification can be made for scenarios with power-law contraction.

Let us now investigate solutions to the mode equation in the form \eqref{eq:ModifiedModeEquation} or \eqref{eq:ModifiedModeEquation2}. First, for large wave vectors $k$ one has $k^2\gg\mathcal{H}^2$ and $k^2\gg |\kappa|$. This corresponds to physical wavelengths $\lambda$ that are small compared to the temporal and spatial curvature scales $H^{-1}$ and $a/\sqrt{|\kappa|}$,
\begin{equation}
    \lambda=a\frac{2\pi}{k}\ll \frac{1}{H}, \quad\quad\quad \lambda = a \frac{2\pi}{k} \ll \frac{a}{\sqrt{|\kappa |}}.
\end{equation}
These short wavelength modes inside the Hubble horizon are standard oscillating modes, as can be seen by solving \eqref{eq:ModifiedModeEquation} neglecting the Hubble and spatial curvature terms, with the two independent solutions $w_k(\eta)\approx e^{-i\abs{k}\eta}$ and $w^*_k(\eta)\approx e^{i\abs{k}\eta}$. Using the normalization condition \eqref{eq:ModeRelation} gives
\begin{equation}
    v_k(t) = \frac{e^{-i\abs{k}\int_{t_0}^t \text{d}t^{\prime} a^{-1}(t^{\prime})}}{\sqrt{2 \abs{k} a(t)}},
\end{equation}
and its complex conjugate as the two solutions for the mode functions in this regime. The solutions describe standard propagating waves on a curved and expanding background.

Another regime that is easily understood arises when $\mathcal{H}$ is small, so that the scale factor $a(t)$ is approximately constant. Here, one finds standard pairs of oscillating modes with the dispersion relation encoded in Eq.\ \eqref{eq:LaplaceEigenvalues}. Note in particular that negative spatial curvature leads to an effective mass $m=\sqrt{|\kappa|}/2$.

Let us now address the regime of small or intermediate wave numbers $k$ compared to the conformal Hubble rate $\mathcal{H}$. This discussion is best done individually for the different expansion scenarios introduced above. We concentrate here on vanishing spatial curvature $\kappa=0$.
\begin{itemize}
\item[(A)] For an accelerated expansion the conformal Hubble rate $\mathcal{H}$ is increasing and a mode with given comoving wave number $k$ will have a transition from ``inside the horizon'' or $k>\mathcal{H}$ to outside the horizon or $k< \mathcal{H}$. Deep inside the horizon one has standard oscillating behavior, while large physics wavelength modes outside the horizon with
\begin{equation}
    \lambda = a \frac{2\pi}{k} \gg \frac{1}{H},
\end{equation}
behave somewhat differently. One finds there for $w_k$ the two independent solutions 
\begin{equation}
|\eta|^{-\frac{\gamma}{2(\gamma-1)}} \quad\quad\text{and} \quad\quad |\eta|^\frac{3\gamma-2}{2(\gamma-1)},
\label{eq:twoSolutionsFarOutsideHorizon}
\end{equation}
corresponding to a growing and decaying behavior. We have used here $\mathcal{H} = \dot a = -\gamma/((\gamma-1)\eta)$ where the conformal time $\eta$ takes negative values evolving towards $\eta=0$. Qualitatively, modes are first oscillating until they ``leave the horizon'' and then evolve algebraically as a superposition of the above two solutions.
\item[(B)] For the coasting universe with $\gamma=1$ and $q=0$ one has far outside the horizon the two independent solutions $e^{Q\eta/2}$ and $e^{-Q\eta/2}$ corresponding to $v_k(t)\approx \text{const.}$ and $v_k(t)\approx a^{-1}(t)$, respectively. Note that for these real solutions the normalization condition in Eq.\ \eqref{eq:ModeRelation} cannot be applied. Because the conformal Hubble rate is constant for $\gamma=1$, every comoving wave number $k$ can be classified as oscillating (when the square bracket in Eq.\ \eqref{eq:ModifiedModeEquation} is positive) or growing/decaying (when the square bracket in Eq.\ \eqref{eq:ModifiedModeEquation} is negative) with respect to the evolution of $w_k(\eta)$ with conformal time $\eta$. Allowing for nonzero spatial curvature $\kappa$ does not change this qualitative picture as long as $|\kappa| < \mathcal{H}^2$. The transition between the two regimes is at the critical comoving wave number
\begin{equation}
k_c= \begin{cases}
    \frac{1}{2} (\sqrt{\mathcal{H}^2+|\kappa|}-\sqrt{|\kappa|})& \text{for}\hspace{0.3cm} \kappa > 0, \\
     \frac{1}{2} \mathcal{H} & \text{for}\hspace{0.3cm} \kappa = 0, \\
    \frac{1}{2}\sqrt{\mathcal{H}^2-|\kappa|} & \text{for}\hspace{0.3cm} \kappa < 0.
    \end{cases}
\end{equation}

In this particular case with $\gamma=1$ modes do not cross the horizon. 

\item[(C)] For decelerated universes the conformal Hubble rate is decreasing, such that modes with a given wave number $k$ enter the horizon at some point in time and start then an oscillating behavior. Modes with very large wavelength that are far outside the horizon still have the two independent solutions in Eq.~\eqref{eq:twoSolutionsFarOutsideHorizon}, but now the conformal time $\eta$ is positive and increasing.
\item[(D)] For the specific case of $\gamma=2/3$ the term in round brackets in \eqref{eq:ModifiedModeEquation} drops out and the solutions are simply plane waves in conformal time, $e^{\pm i |k| \eta}$.
\item[(E)] For a decelerated expansion with $\gamma<2/3$ the ``effective mass squared'' proportional $\mathcal{H}^2$ is positive and decreasing in time, such that one expects some kind of oscillating solution for all wave numbers $k$. 
\item[(F)] Finally, for $\gamma=0$ one has vanishing conformal Hubble rate and standard oscillating behavior for all wave numbers.
\end{itemize} 
Let us mention here that one can actually find analytic solutions for many choices of $\gamma$, some of which are presented in the Appendix.\\

\begin{figure*}[t!]
\centering
\includegraphics[width=0.95\textwidth]{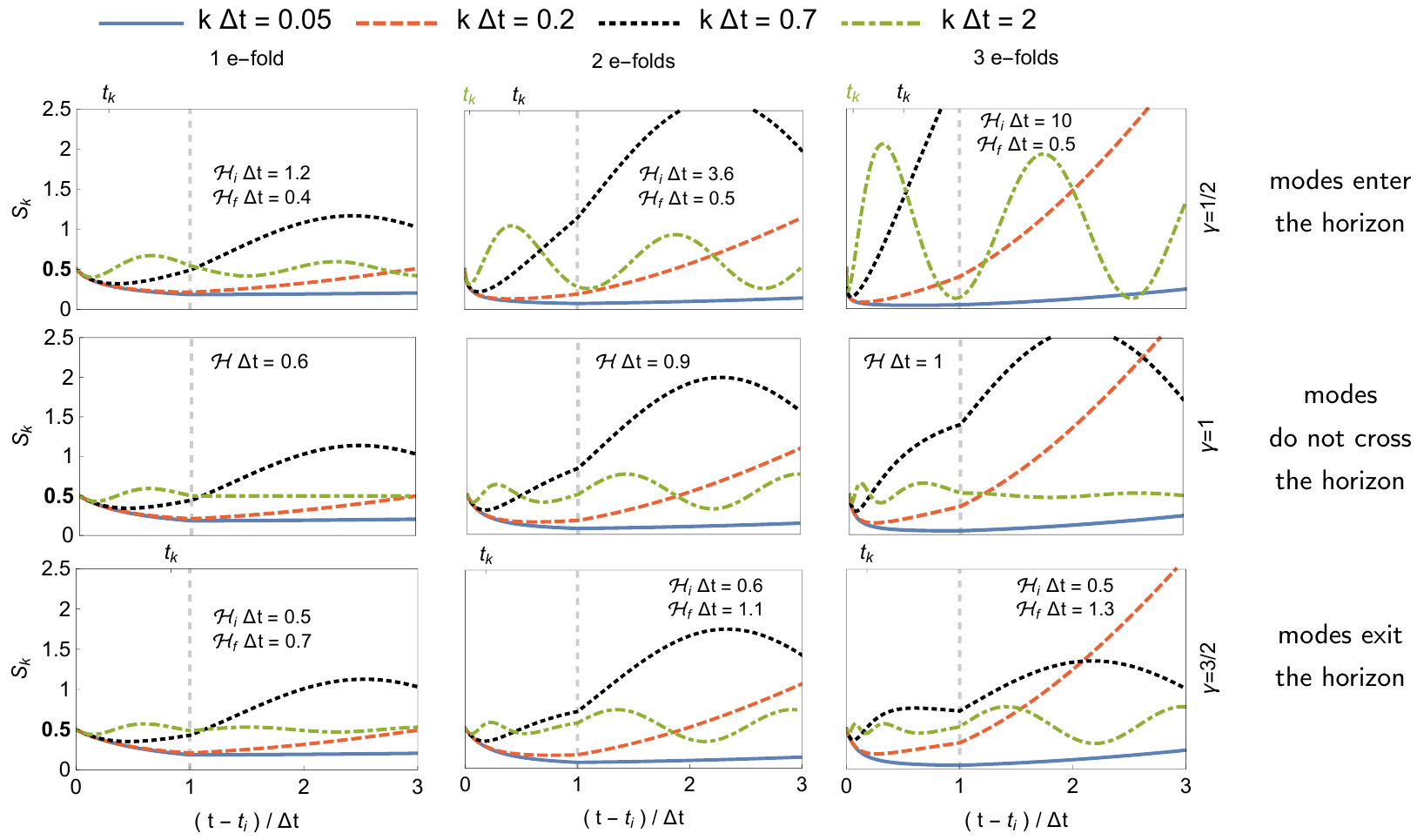}
\caption{Time evolution of certain modes during and after expansion, for a decelerating ($\gamma = 1/2$), coasting ($\gamma = 1$), and accelerating ($\gamma = 3/2$) universe. We indicate the initial and final value of the (conformal) Hubble horizon $\mathcal{H} = \dot{a}$ for each different scenario, for three different e-folds. In the particular case of $\gamma = 1/2$ modes enter the horizon through time. Which modes enter ($k = 0.7 \Delta t^{-1}$ and $k=2 \Delta t^{-1}$) and at which point in time is indicated in upper ticks, with corresponding color (online). In a coasting universe all modes remain either inside or outside the horizon, while for the accelerating universe modes (in this case only $k = 0.7 \Delta t^{-1}$) exit the horizon. In some cases their oscillating or decreasing/increasing nature can be appreciated, corresponding to modes that are inside or outside the horizon. In particular, the two lowest values of $k$ remain outside the horizon for all given scenarios. The horizon disappears when expansion ceases and therefore all modes become oscillatory.}
\label{fig:SpectrumEvolution}
\end{figure*}

\begin{figure*}[t!]
\centering
\includegraphics[width=0.95\textwidth]{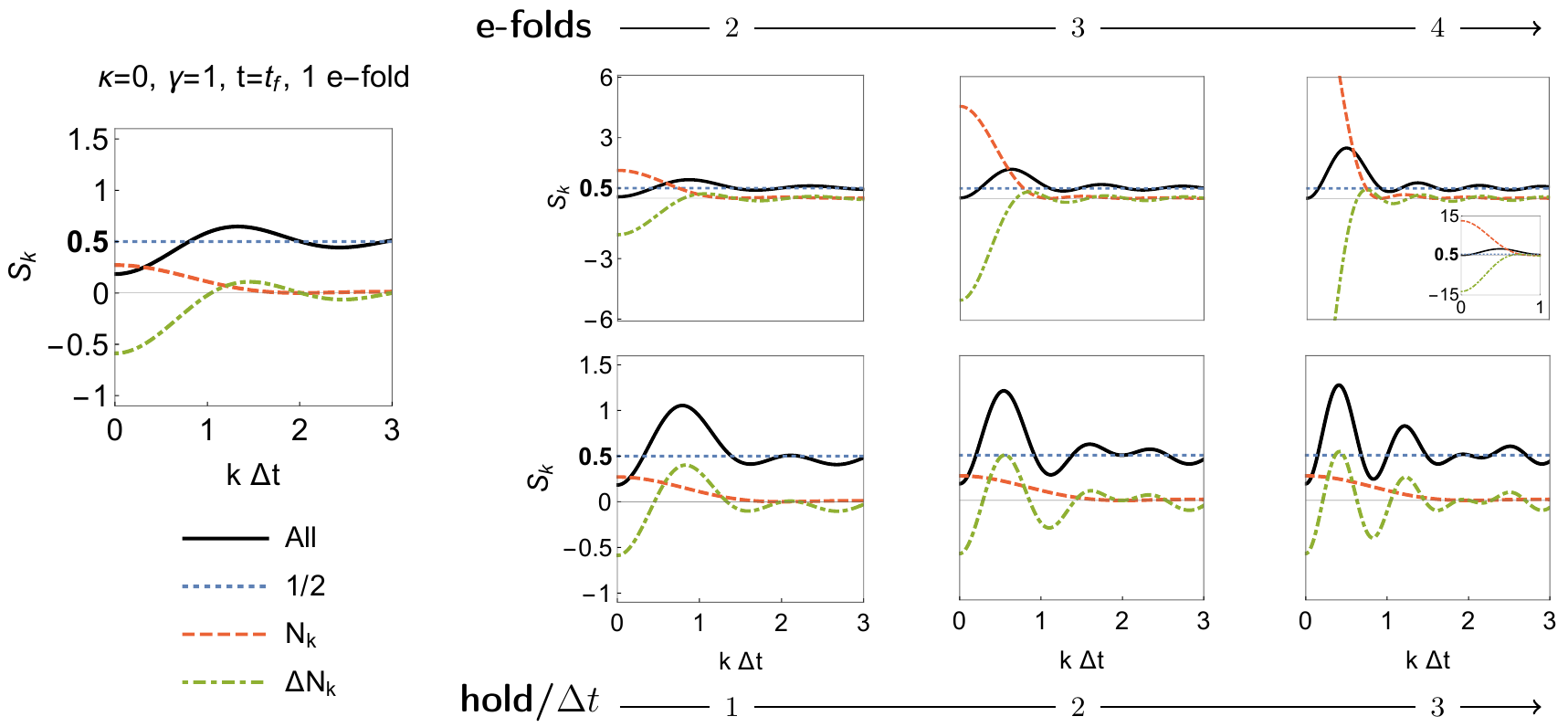}
\caption{Different contributions to the power spectrum $S_k$ as a function of the radial wave number $k$. Left column: Contributions to the spectrum right at the end of expansion. It is noticeable that the time-dependent term $\Delta N_k$ has a larger weight than $N_k$. Right column: In the upper row, the spectrum is plotted for different e-fold numbers, whereas in the lower row, it is done so at different hold times after the expansion has ceased. We observe that increasing $n$ leads to enhanced particle production. The hold time behavior, dependent exclusively on $\Delta N_k$, shows how the power is distributed through time, favoring the lower momentum modes.}
\label{fig:Spectrum}
\end{figure*}

\subsection{Spectrum and two-point correlation functions}
\label{subsec:SpectrumAndCorrelators}

The phenomenon of particle production due to a time-dependent metric can be analyzed in terms of two point functions in position space and their corresponding power spectrum in momentum space. To look into these quantities we place ourselves in a static situation after expansion, i.e. at times $t \ge t_{\text{f}}$ in Eq.\ \eqref{eq:ScaleFactorPolynomial}, and consider equal-time two-point correlation functions of fields at two positions $(u, \varphi)$ and $(u^\prime, \varphi^\prime)$. As a consequence of statistical homogeneity and isotropy, all two-point correlation functions depend on spatial coordinates only through the (comoving) distance $L$, given by
\begin{widetext}
    \begin{equation}
        L = \begin{cases}
        \frac{1}{\sqrt{\abs{\kappa}}}\cos^{-1}\left(\cos\theta\cos\theta^{\prime} + \sin\theta \sin\theta^{\prime} \cos(\varphi - \varphi^{\prime})\right) &\text{for} \hspace{0.1cm} \kappa > 0, \\
        \left[u^2+u^{\prime 2}-2uu^{\prime}\cos(\varphi-\varphi^{\prime})\right]^{1/2} &\text{for} \hspace{0.1cm} \kappa = 0, \\
        \frac{1}{\sqrt{\abs{\kappa}}} \cosh^{-1}\left(\cosh \sigma \cosh \sigma^{\prime} - \sinh \sigma \sinh \sigma^{\prime} \cos(\varphi-\varphi^{\prime})\right) &\text{for} \hspace{0.1cm} \kappa < 0.
        \end{cases}
    \end{equation}
\end{widetext}
This property of the correlators is a consequence of the symmetries of the FLRW universe.

The statistical equal-time two-point correlation function of the field $\phi$ is then
\begin{equation}
    \begin{split}
        \mathcal{G}_{\phi \phi} (t,L) 
        & = \frac{1}{2}\braket{\{{\phi}(t,u,\varphi), {\phi}(t,u^{\prime},\varphi^{\prime})\}}_c \\
        &=\int_k \, \mathcal{F}(k, L) \frac{1}{a_\text{f} \sqrt{-h(k)}} \tilde{S}_k(t),
        \end{split}
    \label{eq:CorrelationFunctionPhiPhiEqualTime}
\end{equation}
with the integral measure in momentum space
\begin{equation}
    \int_{k}  = \begin{cases}
    \sum_{l=0}^{\infty} \abs{\kappa}\frac{l+1/2}{2\pi} &\text{for} \hspace{0.2cm} \kappa > 0, \\
    \int \frac{\text{d}k}{2\pi} \, k  &\text{for} \hspace{0.2cm} \kappa = 0,\\
    \int \frac{\text{d}l}{2\pi} \,\abs{\kappa} l \tanh(\pi l) &\text{for} \hspace{0.2cm} \kappa < 0,
    \end{cases}
\end{equation}
and the integration kernels
\begin{equation}
    \hspace{-0.15cm}\mathcal{F}(k, L) = \begin{cases}
    P_l\left(\cos{\left(L \sqrt{\abs{\kappa}}\right)}\right) &\text{for} \hspace{0.2cm} \kappa > 0, \\
     J_0\left(kL\right) &\text{for} \hspace{0.2cm} \kappa = 0, \\
    P_{il - 1/2}\left(\cosh{\left(L \sqrt{\abs{\kappa}}\right)}\right) &\text{for} \hspace{0.2cm} \kappa < 0.
    \end{cases}
    \label{eq:IntegrationKernels}
\end{equation}
We have introduced the object
\begin{equation}
    \tilde{S}_k(t) = \frac{1}{2} + N_k(t) - \Delta N_k(t),
\end{equation}
determined through the Bogoliubov coefficients discussed in Sec.\ \ref{subsec:BogoliubovCoefficients}, where
\begin{equation}
   N_{k} = \bra{\Omega}\hat{b}^{\dagger}_{km} \, \hat{b}_{km} \ket{\Omega} = |\beta_k|^2,
\label{eq:ParticleNk}
\end{equation}
and
\begin{equation}
    \Delta N_k(t) = \text{Re}\left[ c_k e^{2i\omega_kt}\right], \quad c_k = \alpha_k\beta_k.
    \label{eq:defDeltaN}
\end{equation}
The above $\tilde{S}_k(t)$ is closely related to the power spectrum $S_k(t)$ appearing in the equal-time statistical correlation function of time derivatives of fields,
\begin{equation}
    \begin{split}
        \mathcal{G}_{\dot\phi \dot\phi} (t, L) 
        &= \frac{1}{2}\braket{\{ \dot \phi(t,u,\varphi), \dot \phi(t,u^{\prime},\varphi^{\prime})\}}_c\\
        &= \int_k \, \mathcal{F}(k, L) \frac{\sqrt{-h(k)}}{a_\text{f}^3 } S_k(t),
    \label{eq:CorrelationFunctionPhidotPhidotEqualTime}   
    \end{split}
\end{equation}
where
\begin{equation}
S_k(t) = \frac{1}{2} + N_k(t) + \Delta N_k(t).
\label{eq:Spectrum}
\end{equation}
This can also be written as
\begin{equation}
    S_k (t) = \frac{1}{2} + N_k + \abs{c_k} \cos \left(\theta_k + 2 \omega_k t \right),
    \label{eq:SktWithPhase}
\end{equation}
which specifies explicitly a phase that each $k$-mode acquires after expansion, 
\begin{equation}
    \theta_k = \text{Arg} (c_k).
\label{eq:PhaseDef}
\end{equation}
Given that the power spectrum $S_k(t)$ encodes the effect of particle production per mode, we focus on a rather detailed analysis of this object, in the case of vanishing spatial curvature. 

We begin with Fig.\ \ref{fig:SpectrumEvolution}, where we show the evolution of $S_k(t)$ for some selected modes during and after expansion. The aim of this figure is to show the shape of different modes which are either inside the Hubble horizon, outside of it, or change region during expansion. The range of modes that change in region is greater for larger e-fold number, and this is also explicitly depicted within.

In Fig.\ \ref{fig:Spectrum} we analyze the shape of the three different contributions in \eqref{eq:Spectrum} to $S_k(t)$, and show them in dependence of the number of e-folds of the expansion (upper row), as well as the hold time (time after expansion, lower row). Here, we concentrate on vanishing spatial curvature and a coasting universe with $\gamma=1$. The oscillations visible after the expansion has ceased (green and black curves in the lower row) can be understood as the analog of Sakharov oscillations in the present context.

\begin{figure}
\centering
\includegraphics[width=0.48\textwidth]{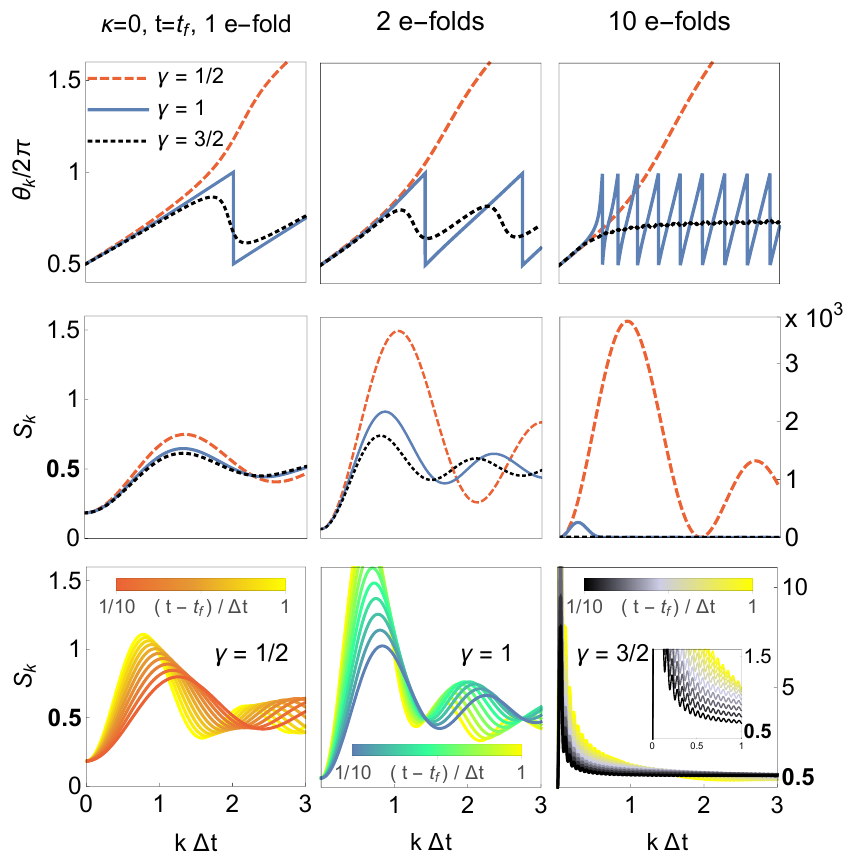}
\caption{Upper row: acquired phase after expansion as a function of wave number $k$ (cf. Eq.\ \eqref{eq:PhaseDef}). The effect of e-fold number is noticeable for coasting and accelerating universes, and has almost no repercussion for a decelerated expansion. Within an accelerating universe the phase difference for different momenta decreases with e-fold number, while its frequency increases. An increase in frequency is also seen for a coasting universe, but the phase difference remains at $\pi$. In particular, for $\gamma = 1$ there are phase jumps at certain momenta $k$ which are not triggered by particle production, $\beta_k = 0$. Middle row: spectra right at the end of expansion. While for low e-fold number the shape of expansion (encoded in $\gamma$) does not affect much the shape of the spectra, this changes when increasing e-folds. In any case, the power spectrum acquires higher values when acceleration is lower, with a big impact for larger e-fold number ($3$ orders of magnitude for $\gamma = 1/2$ when e-fold number goes from $2$ to $10$). Another thing to note is that the peak of the spectrum remains roughly at the same place for a decelerating situation when increasing e-folds, and drifts to lower momentum for both, a coasting and an accelerated universe. Bottom row: evolution of the spectra after expansion has ceased, for a chosen $\gamma$ for each e-fold number, depicting in this way an overview of the situation. The value of the spectrum evolves in time with a frequency set by the dispersion relation $\omega_k = k$, and an initial phase given in the upper row. When $\gamma = 1$ one obtains nodes at $\beta_k = 0$, related to the phase jumps in the upper row. When $\gamma = 3/2$ one can appreciate the highly oscillatory nature of the spectrum (inset in lower right corner) related to highly oscillating phases.}
\label{fig:PhaseAndHold}
\end{figure}

In Fig.\ \ref{fig:PhaseAndHold} we turn first to the phase that each $k$-mode acquires after expansion, defined in Eq.\ \eqref{eq:PhaseDef}, for different  types of expansion (decelerating, coasting, accelerating) and different numbers of e-folds. As seen in Eq.\  \eqref{eq:SktWithPhase} these phases enter the spectrum through $\Delta N_k (t)$ in Eq.\ \eqref{eq:defDeltaN}, and also determine the starting point for the oscillatory evolution of the spectrum after the end of expansion. In the middle and lower row of \autoref{fig:PhaseAndHold} we also discuss $S_k(t)$ right when the expansion has ceased ($t=t_\text{f}$) and its time evolution in a static situation after the expansion. The details of the corresponding analytical calculations can be found in the Appendix.

Additionally, for the mixed statistical equal-time correlation function one has
\begin{equation}
    \begin{split}
    \mathcal{G}_{\phi \dot\phi} (t, L) &=\mathcal{G}_{\dot\phi \phi} (t, L) \\
        &= \frac{1}{2}\braket{ \{\phi(t,u,\varphi) , \dot\phi(t,u^{\prime},\varphi^{\prime})\}}_c \\
        &=\int_k \mathcal{F}(k,L) \frac{1}{a_{\text{f}}^2} \text{Im}\left[ c_k e^{2i\omega_kt}\right].
    \end{split}
    \label{eq:CorrelationFunctionPhiPhidotEqualTime}
\end{equation}

As they stand, the three two-point correlation functions in Eqs.\ \eqref{eq:CorrelationFunctionPhiPhiEqualTime}, \eqref{eq:CorrelationFunctionPhidotPhidotEqualTime} and \eqref{eq:CorrelationFunctionPhiPhidotEqualTime} show ultraviolet divergences. These can be cured through the use of test or window functions, which act as a regulator. The correlation function of ``smeared out'' fields becomes
\begin{equation}
\begin{split}
  G(t,L) 
  =&\, \frac{a_{\text{f}}}{m} \int_k \mathcal{F}(k,L) T(t,k) \tilde W^*(k) \tilde W(k),
\end{split}
    \label{eq:ConvolutedCorrelationFunction}
\end{equation}
where $T(t,k)$ specifies the correlation function or spectrum in momentum space before regularization, and $\tilde W(k)$ denotes the window function in momentum space. With Eq.\ \eqref{eq:ConvolutedCorrelationFunction} at hand, all two-point correlation functions can be calculated. 

\begin{figure*}
\centering
\includegraphics[width=0.95\textwidth]{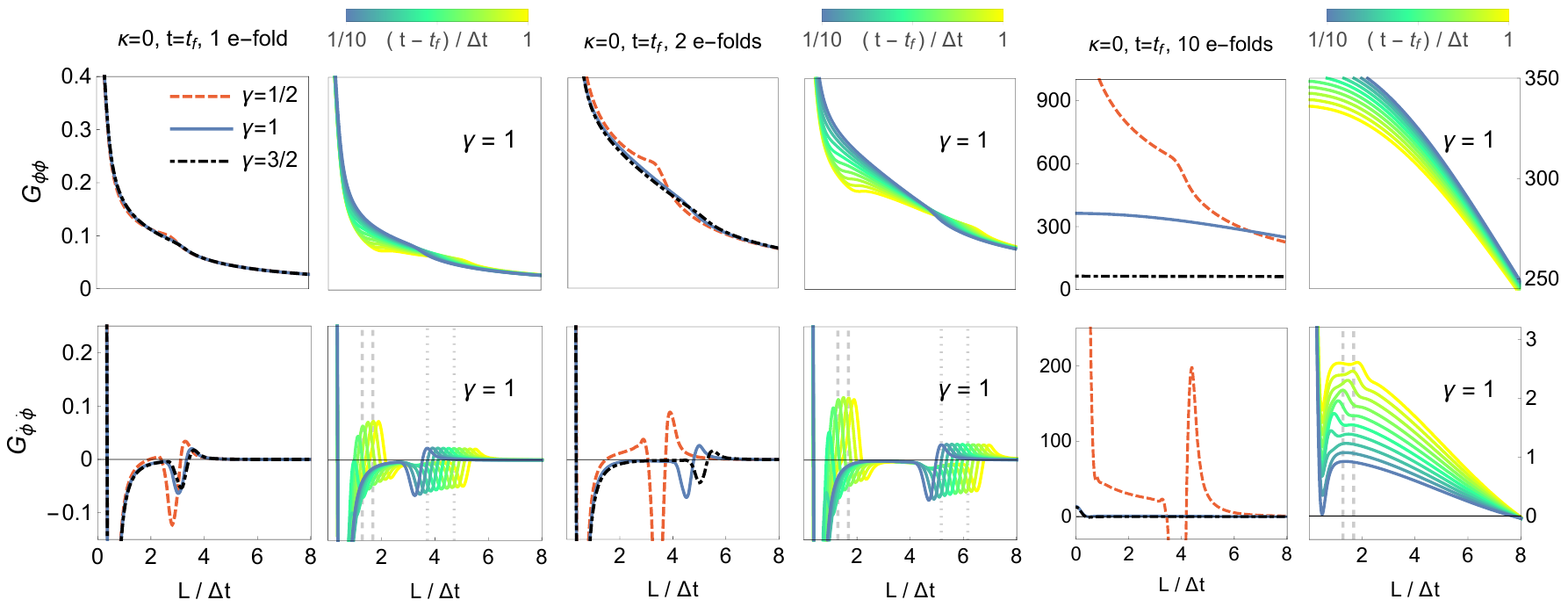}
\caption{Upper row: Regularized field-field correlation function for a decelerating ($\gamma = 1/2$), uniform ($\gamma = 1$), and accelerating ($\gamma = 3/2$) universe at the time where expansion ceases, and its evolution in a static situation after uniform expansion, for three different numbers of e-folds. Lower row: Same analysis but for time derivatives of fields. Depending on $\gamma$, a change in the number of e-folds has a different effect. For example, the shape of the correlation function is roughly the same for $\gamma = 1/2$ at different numbers of e-folds. In contrast, for $\gamma \geq 1$ correlations are ``elongated'' in space. Additionally, for $\gamma \leq 1$, an increase in e-fold number enhances correlations, while for an accelerating universe, the magnitude of correlations seems to decrease with e-fold number. Finally, an important feature seen in the time evolution plots is the propagation of correlations at twice the speed of light $c=1$, indicated through vertical lines. We use dashed vertical lines to indicate peaks which are separated by $0.4 c$, and dotted ones for peaks separated by $c$. 
}
\label{fig:TwoPointFunctions}
\end{figure*}

In Fig.\ \ref{fig:TwoPointFunctions} we show field-field and time derivative of fields correlation functions right at the end of expansion and their evolution after expansion has ceased, for a Gaussian window function of width $w^2 = 0.02 \,\Delta t^2$. For a spatially flat scenario, this translates to a regularization in momentum space by 
\begin{equation}
   \tilde W^*(k)\tilde W(k) =e^{- w^2 k^2}.
\end{equation}
While the position of correlation peaks in the two-point functions is a robust feature with respect to the choice of $w$, their magnitude and range is not. This becomes especially important for larger number of e-folds.

In the position space representation of \autoref{fig:TwoPointFunctions}, particle production is visible as combination of correlations and anti-correlations at intermediate distances, which, for a massless field, move to larger distances with the velocity of light after the expansion has stopped. The particular shape of the correlation function necessarily depends somewhat on the form of the test function.

\section{Conclusion and Outlook}
\label{sec:Outlook}
Motivated by the application to Bose-Einstein condensates as quantum simulators, we have discussed mode equations and particle production for a massless relativistic scalar field in $d=2+1$ dimensional cosmologies. Our results are mainly analytical and encompass positive, vanishing and negative spatial curvature. Depending on the form of the expansion, which can be accelerating, coasting or decelerating, and the number of e-folds, this leads to a rather rich phenomenology of observables in terms of different power spectra and two-point correlation functions. We have concentrated here in particular on equal-time correlation functions of fields and their time derivatives or conjugate momenta.

In the future, it could be interesting to investigate also composite operators such as the energy-momentum tensor and its renormalization, and to extend the analysis to other types of matter fields, such as fermions or gauge fields. Also the physics of (cosmological) spacetime horizons and related phenomena involving the dynamics of entanglement are worth further study.

Of particular interest is also investigation of the physics discussed here with quantum simulators as they can be realized, e.g., with ultracold atoms (see also our companion papers \cite{BECPaper2022} and \cite{ExperimentPaper2022}). Furthermore, one may investigate other nontrivial spacetime geometries in $d=2+1$ spacetime dimensions and try to establish one-to-one correspondences to table-top experiments. In this way, effects of quantum fields in curved spacetime become accessible in the laboratory.  

\section*{Acknowledgements}
The authors would like to thank Celia Viermann, Marius Sparn, Nikolas Liebster, Maurus Hans, Elinor Kath, Helmut Strobel, Markus Oberthaler, Finn Schmutte and Simon Brunner for fruitful discussions. This work is supported by the Deutsche Forschungsgemeinschaft (DFG, German Research Foundation) under Germany's Excellence Strategy EXC 2181/1 - 390900948 (the Heidelberg STRUCTURES Excellence Cluster) and under SFB 1225 ISOQUANT - 273811115 as well as FL 736/3-1. NSK is supported by the Deutscher Akademischer Austauschdienst (DAAD, German Academic Exchange Service) under the Länderbezogenes Kooperationsprogramm mit Mexiko: CONACYT Promotion, 2018 (57437340). APL is supported by the MIU (Spain) fellowship FPU20/05603 and the MICINN (Spain) project PID2019-107394GB-I00 (AEI/FEDER, UE).



\appendix

\section{Mode functions and Bogoliubov coefficients for different expansion histories}
\label{app:PPSpectra}

In this appendix we collect analytic expressions for mode functions and Bogoliubov coefficients for different cosmological expansion histories. A distinction is made between three temporal regions,
\begin{itemize}
    \item Region I: early times $t \le t_\text{i}$, constant scale factor $a=a_\text{i}$,
    \item Region II: intermediate times $t_\text{i}<t<t_\text{f}$, evolving scale factor $a=a(t)$,
    \item Region III: late times $t_\text{f} \le t$, constant scale factor $a=a_\text{f}$.
\end{itemize}
Expansion histories differ by the time dependent scale factor $a(t)$ in the intermediate region II. Note that the following expressions are general for all three types of spatial curvature $\kappa$ and hence are given in terms of $h(k)$.

\subsection{Power-law expansion with exponent $\gamma=1/2$}
\label{subsec:powerLawgamma12}

We start with a power-law form of the scale factor as a functions of time corresponding to $\gamma = 1/2$ in \eqref{eq:ScaleFactorPolynomial}, setting $t_0 = 0$
\begin{equation}
    a(t) = Q\sqrt{\abs{t}},
\end{equation}
and focus on the solutions to the mode equation \eqref{eq:ModeEquation} for the three regions.

\paragraph*{Region I.}

In Region I, the general solution for the mode equation \eqref{eq:ModeEquation} is a superposition of plane waves with positive and negative frequencies,
\begin{equation}
    v_k^{\text{I}}(t) = A_k^{\text{I}} e^{-i\omega_k^{\text{I}}t} + B_k^{\text{I}} e^{i\omega_k^{\text{I}}t},
\label{eq:RegionIPlaneWave}
\end{equation}
where $A_k^{\text{I}}$ and $B_k^{\text{I}}$ are some coefficients and the frequency is given by
\begin{equation}
    \omega_k^{\text{I}} = \frac{\sqrt{-h(k)}}{a_\text{i}} = \sqrt{\frac{-h(k)}{Q^2t_\text{i}}}.
\end{equation}
We define the $a$-particles as corresponding to the positive frequency modes, such that $B_k^{\text{I}} = 0$ and
\begin{equation}
    v_k^{\text{I}}(t)=\frac{e^{-i\omega_k^{\text{I}}t}}{a_\text{i}\sqrt{2 \omega_k^{\text{I}}}},
\label{eq:SolutionModeEquationRegionI}
\end{equation}
where $A_{k}^{\text{I}} = 1/\left(a_\text{i}\sqrt{2 \omega_{k}^{\text{I}}}\right)$ is a consequence of the normalization condition \eqref{eq:ModeRelation}, which in case of plane waves reduces to
\begin{equation}
    \omega_k^{\text{I}} \left(\abs{A_k^{\text{I}}}^2 - \abs{B_k^{\text{I}}}^2 \right) = \frac{1}{2 a^2(t)}.
\label{eq:BogoliubovRelationPlaneWaves}
\end{equation}

\paragraph*{Region II.}
Now, the general solution of the mode equation \eqref{eq:ModeEquation} is given by
\begin{equation}
    v_k^{\text{II}}(t) = A_k^{\text{II}} J_0\left(2\frac{\sqrt{-h(k)t}}{Q}\right) +B_k^{\text{II}} Y_0 \left(2\frac{\sqrt{-h(k)t}}{Q}\right),
    \label{eq:ModeFunctionRegionII}
\end{equation}
where $J_0$ and $Y_0$ are Bessel functions of the first and second kind, respectively. The coefficients $A_k^{\text{II}}$ and $B_k^{\text{II}}$ are determined by requiring that the mode function and its first derivative in region II match with those in region I,  at their joint boundary, i.e.
\begin{equation}
    v_k^{\text{I}}(t_\text{i}) \overset{!}{=} v_k^{\text{II}}(t_\text{i}), \quad\quad\quad \dot{v}_k^{\text{I}}(t_\text{i}) \overset{!}{=} \dot{v}_k^{\text{II}}(t_\text{i}).
\label{eq:ModeFunctionsJointBoundaryi}
\end{equation}
Then, the coefficients evaluate to
\begin{equation}
    \begin{split}
        &\hspace{-0.15cm}A_k^{\text{II}} = \frac{ \pi\sqrt{\omega^{\text{I}}_kt_\text{i}}}{\sqrt{2}Q} e^{-i\omega_k^{\text{I}}t_\text{i}} 
        \Big[i Y_0\left(2\omega_k^{\text{I}}t_\text{i}\right) - Y_1\left(2\omega_k^{\text{I}}t_\text{i}\right)\Big],\\
        &\hspace{-0.15cm}B_k^{\text{II}} = - \frac{ \pi \sqrt{\omega^{\text{I}}_kt_\text{i}}}{\sqrt{2}Q} e^{-i\omega_k^{\text{I}}t_\text{i}}
        \Big[i J_0\left(2\omega_k^{\text{I}}t_\text{i}\right)  - J_1\left(2\omega_k^{\text{I}}t_\text{i}\right)\Big],
    \end{split}
    \label{eq:CoefficientsRegionIIProperPhase}
\end{equation}
where $J_1$ and $Y_1$ are again Bessel functions of the first and second kind.

\paragraph*{Region III.}
As for Region I, the solution is of plane wave form
\begin{equation}
    v_k^{\text{III}}(t) = A_k^{\text{III}} e^{-i\omega_k^{\text{III}}t} + B_k^{\text{III}} e^{i\omega_k^{\text{III}}t},
\label{eq:planeWaveRegionIIIgamma1/2}
\end{equation}
but the frequencies in general differ
\begin{equation}
    \omega_k^{\text{III}} =  \frac{\sqrt{-h(k)}}{a_\text{f}} = \sqrt{\frac{-h(k)}{Q^2 t_\text{f}}}\neq \omega_k^{\text{I}},    
\end{equation}
and the new coefficients $A_k^{\text{III}}$ and $B_k^{\text{III}}$ have to be determined by comparing to the mode function in region II and its first derivative at the shared boundary
\begin{equation}
    v_k^{\text{II}}(t_\text{f}) \overset{!}{=} v_k^{\text{III}}(t_\text{f}), \quad\quad\quad \dot{v}_k^{\text{II}}(t_\text{f}) \overset{!}{=} \dot{v}_k^{\text{III}}(t_\text{f}).
\label{eq:BoundaryConditionf}
\end{equation}
Then we find
\begin{widetext}
\begin{equation}
    \begin{split}
        A_k^{\text{III}} =  \frac{\pi\sqrt{\omega^{\text{I}}_kt_\text{i}}}{2\sqrt{2}Q} e^{i(\omega^{\text{III}}_k t_\text{f} - \omega^{\text{I}}_k t_\text{i})}  &\Biggr{\{} \left[ i Y_0\left(2\omega^{\text{I}}_kt_\text{i}\right)-Y_1\left(2\omega^{\text{I}}_kt_\text{i}\right) \right] \left[J_0\left(2\omega^{\text{III}}_k t_\text{f}\right)- i J_1 \left(2\omega^{\text{III}}_k t_\text{f}\right)\right]\\
        &-\left[iJ_0\left(2\omega^{\text{I}}_kt_\text{i} \right)-J_1\left(2\omega^{\text{I}}_kt_\text{i}\right)\right]\left[Y_0 \left(2\omega^{\text{III}}_k t_\text{f}\right)-iY_1\left(2\omega^{\text{III}}_k t_\text{f}\right)\right] \Biggr{\}},
    \end{split}
\end{equation}
and
\begin{equation}
    \begin{split}
        B_k^{\text{III}} = \frac{\pi\sqrt{\omega^{\text{I}}_kt_\text{i}}}{2\sqrt{2}Q} e^{-i(\omega^{\text{III}}_k t_\text{f} + \omega^{\text{I}}_k t_\text{i})}  &\Biggr{\{} \left[ i Y_0\left(2\omega^{\text{I}}_kt_\text{i}\right)-Y_1\left(2\omega^{\text{I}}_kt_\text{i}\right) \right] \left[J_0\left(2\omega^{\text{III}}_k t_\text{f}\right)+i J_1 \left(2\omega^{\text{III}}_k t_\text{f}\right)\right]\\
        &-\left[iJ_0\left(2\omega^{\text{I}}_kt_\text{i} \right)-J_1\left(2\omega^{\text{I}}_kt_\text{i}\right)\right]\left[Y_0 \left(2\omega^{\text{III}}_k t_\text{f}\right)+iY_1\left(2\omega^{\text{III}}_k t_\text{f}\right)\right] \Biggr{\}}.
    \label{eq:CoefficientsRegionIIIProperPhase}
    \end{split}
\end{equation}
\end{widetext}
Since in region III we again have the Minkowskian vacuum and plane wave solutions (with positive and negative frequency) for the mode functions, we can write \eqref{eq:planeWaveRegionIIIgamma1/2} as
\begin{equation}
\begin{split}
   v_k^{\text{III}}(t) &= a_\text{f} \sqrt{2 \omega_k^{\text{III}}} \left[ A_k^{\text{III}} u_k (t) + B_k^{\text{III}}  u_k^{*} (t) \right],
    \label{eq:FLRWModeFunctionIIIvu}
\end{split}
\end{equation}
where 
\begin{equation}
     u_k(t) \equiv \frac{e^{-i\omega_k^{\text{III}}t} }{a_\text{f}\sqrt{2 \omega_k^{\text{III}}}},
\label{eq:ModeFunctionUIII}
\end{equation}
are the positive frequency plane waves in region III. Then, using the inverse Bogoliubov relation between $v_k$ and $u_k$ given in Eq.\  \eqref{eq:BogoliubovTransformationOperators} 
\begin{equation}
    v_k (t) = \alpha_k^* u_k (t) - \beta_k u_k^{*} (t)
\end{equation}
and comparing with \eqref{eq:FLRWModeFunctionIIIvu} we can identify
\begin{equation}
    \alpha^*_k = A_k^{\text{III}} a_{\text{f}} \sqrt{2 \omega_k^{\text{III}}}, \quad  \beta_k = -B_k^{\text{III}} a_\text{f} \sqrt{2 \omega_k^{\text{III}}}.   
\end{equation}
All further results can be deduced from the above as discussed around Eq.\ \eqref{eq:Spectrum}

\subsection{Power-law expansion with exponent $\gamma = 2/3$}

We study now the case $\gamma = 2/3$ so that the scale factor in region II is
\begin{equation}
    a(t) = Q\abs{t}^{2/3}.
\end{equation}

\paragraph*{Region I.} Before the expansion the situation is equivalent to the one discussed in Sec.\ \ref{subsec:powerLawgamma12}, so that the relations given there are also the starting point here.

\paragraph*{Region II.} For this case, the general solution is
\begin{align}
    v_{k}^{\text{II}}(t) = A_{k}^{\text{II}}\,\frac{e^{-3 i \omega^{\text{II}}_{k} t}}{t^{1/3}} + B_{k}^{\text{II}}\,\frac{e^{3 i \omega^{\text{II}}_{k} t}}{t^{1/3}}.
    \label{eq:ModeFunctionRegionIIGamma2/3}
\end{align}
We determine as before the coefficients $A_{k}^{\text{II}}$ and $B_{k}^{\text{II}}$ from the matching condition in Eq.\ \eqref{eq:ModeFunctionsJointBoundaryi} and obtain
\begin{equation}
    \begin{split}
        &A_{k}^{\text{II}} =\left( \frac{1}{\sqrt{2 Q \sqrt{-h(k)}}} + \frac{i \sqrt{Q}}{6 t_\text{i}^{1/3}\sqrt{2 (-h(k))^{3/2}}} \right)e^{2 i \omega^{\text{I}}_{k} t_\text{i}}
    \end{split}
\end{equation}
and
\begin{equation}
    \begin{split}
        &B_{k}^{\text{{II}}}  =\left( \frac{- i \sqrt{Q}}{6 t_\text{i}^{1/3}\sqrt{2 (-h(k))^{3/2}}} \right)e^{-4 i \omega^{\text{I}}_{k} t_\text{i}}.
    \end{split}
\end{equation}

\paragraph*{Region III.} We have once more the general solution
\begin{equation}
    v_{k}^{\text{III}}(t) = A_{k}^{\text{III}} e^{-i\omega_{k}^{\text{III}}t} + B_{k}^{\text{III}} e^{i\omega_{k}^{\text{III}}t},
\label{eq:planeWaveRegionIIIgamma2/3}
\end{equation}
with frequency
\begin{equation}
    \omega_{k}^{\text{III}} =  \frac{\sqrt{-h(k)}}{a_\text{f}} = \frac{\sqrt{-h(k)}}{Qt_\text{f}^{2/3}}.    
\end{equation}
The boundary matching condition \eqref{eq:BoundaryConditionf} allows to solve for the coefficients, as

\begin{widetext}
\begin{equation}
    \begin{split}
        A_{k}^{\text{III}} &=  \frac{1} {2}  \left(\frac{2}{t_\text{f}^{1/3}}- \frac{i Q}{3\sqrt{-h(k)}\,t_\text{f}^{2/3}}
       \right)\left(\frac{1}{\sqrt{2Q\sqrt{-h(k)}}} + \frac{i\sqrt{Q}}{3\left(2\sqrt{-h(k)}\right)^{3/2}t_\text{i}^{1/3}} \right) e^{-2i (\omega_{k}^{\text{III}}t_\text{f}-\omega_{k}^{\text{I}}t_\text{i})} \\
        &\hspace{1cm}+ \frac{1} {2}  \left( \frac{i Q}{3\sqrt{-h(k)}\,t_\text{f}^{2/3}}
       \right)\left( \frac{i\sqrt{Q}}{3\left(2\sqrt{-h(k)}\right)^{3/2}t_\text{i}^{1/3}} \right) e^{4i (\omega_{k}^{\text{III}}t_\text{f}-\omega_{k}^{\text{I}}t_\text{i})},
    \end{split}
\end{equation}
and
\begin{equation}
    \begin{split}
        B_{k}^{\text{III}} &= \frac{1} {2}  \left( \frac{i Q}{3\sqrt{-h(k)}\,t_\text{f}^{2/3}}
       \right)\left(\frac{1}{\sqrt{2Q\sqrt{-h(k)}}} + \frac{i\sqrt{Q}}{3\left(2\sqrt{-h(k)}\right)^{3/2}t_\text{i}^{1/3}} \right) e^{-2i (2\omega_{k}^{\text{III}}t_\text{f}-\omega_{k}^{\text{I}}t_\text{i})} \\
        &\hspace{1cm}+ \frac{1} {2}  \left( \frac{2}{t_\text{f}^{1/3}}+\frac{i Q}{3\sqrt{-h(k)}\,t_\text{f}^{2/3}}
       \right)\left( \frac{-i\sqrt{Q}}{3\left(2\sqrt{-h(k)}\right)^{3/2}t_\text{i}^{1/3}} \right) e^{2i (\omega_{k}^{\text{III}}t_\text{f}-2\omega_{k}^{\text{I}}t_\text{i})}.
\end{split}
\end{equation}
\end{widetext}
We can then write \eqref{eq:planeWaveRegionIIIgamma2/3} as in equation \eqref{eq:FLRWModeFunctionIIIvu}, in terms of the positive frequency plane waves in region III \eqref{eq:ModeFunctionUIII}.

\subsection{Power-law expansion with exponent $\gamma = 1$}

We study now the case $\gamma = 1$ so that the scale factor in region II is
\begin{equation}
    a(t) = Q\abs{t}.
\end{equation}

\paragraph*{Region I.} Before the expansion the situation is analogous to the one discussed in Sec.\ \ref{subsec:powerLawgamma12} and the relations given there can be easily transferred.

\paragraph*{Region II.} For this case, the general solution is
\begin{equation}
\begin{split}
    v_{k}^{\text{II}}(t) &= A_{k}^{\text{II}} t^{-\frac{1}{2}\sqrt{1-4\Omega_1}-\frac{1}{2}}+ B_{k}^{\text{II}}t^{\frac{1}{2}\sqrt{1-4\Omega_1}-\frac{1}{2}},
\label{eq:ModeFunctionRegionIIOpen}
\end{split}
\end{equation}
where we introduced the $k$-dependent variable ${\Omega_{1}=-h(k)/Q^2}$. 

We determine as before the coefficients $A_{k}^{\text{II}}$ and $B_{k}^{\text{II}}$ from the matching condition in Eq.\ \eqref{eq:ModeFunctionsJointBoundaryi}, leading to
\begin{equation}
    \begin{split}
        A_{k}^{\text{II}} &=\frac{1}{Q\sqrt{2 \omega_k^{\text{I}}}} \frac{e^{-i\omega_{k}^{\text{I}}t_\text{i}}}{\sqrt{1-4\Omega_1}}t_\text{i}^{\frac{1}{2}\sqrt{1-4\Omega_1}+\frac{1}{2}}\\
        &\hspace{1cm}\times\Big[i\omega_{k}^{\text{I}} + \frac{1}{2t_\text{i}}\left(\sqrt{1-4\Omega_1}-1\right)\Big]
    \end{split}
    \label{eq:CoefficientsRegionIIProperPhaseOpenA}
\end{equation}
and
\begin{equation}
    \begin{split}
        B_{k}^{\text{{II}}} &= -\frac{1}{Q\sqrt{2 \omega_k^{\text{I}}}} \frac{e^{-i\omega_{k}^{\text{I}}t_\text{i}}}{\sqrt{1-4\Omega_1}}t_\text{i}^{-\frac{1}{2}\sqrt{1-4\Omega_1}+\frac{1}{2}}\\
        &\hspace{1cm}\times\Big[i\omega_{k}^{\text{I}} - \frac{1}{2t_\text{i}}\left(\sqrt{1-4\Omega_1}+1\right)\Big].
    \end{split}
    \label{eq:CoefficientsRegionIIProperPhaseOpenB}
\end{equation}
\vspace{1.1cm}
\paragraph*{Region III.} Again, we have the general solution
\begin{equation}
    v_{k}^{\text{III}}(t) = A_{k}^{\text{III}} e^{-i\omega_{k}^{\text{III}}t} + B_{k}^{\text{III}} e^{i\omega_{k}^{\text{III}}t},
\label{eq:planeWaveRegionIIIOpen}
\end{equation}
with frequency
\begin{equation}
    \omega_{k}^{\text{III}} =  \frac{\sqrt{-h(k)}}{a_\text{f}} = \frac{\sqrt{-h(k)}}{Qt_\text{f}}.    
\end{equation}
Matching at the boundaries, with Eq.\ \eqref{eq:BoundaryConditionf}, one finds
\begin{widetext}
\begin{equation}
    \begin{split}
        A_{k}^{\text{III}} = \frac{1}{Q\sqrt{2 \omega_k^{\text{I}}}} \frac{e^{i(\omega^{\text{III}}_{k}t_\text{f} - \omega^{\text{I}}_{k}t_\text{i})}}{2i\omega^{\text{III}}_{k}\sqrt{1-4\Omega_1}} &\Bigg\{\left[i\omega_{k}^{\text{I}} + \frac{1}{2t_\text{i}}\left(\sqrt{1-4\Omega_1}-1\right)\right]\left[i\omega_{k}^{\text{III}} + \frac{1}{2t_\text{f}}\left(\sqrt{1-4\Omega_1}+1\right)\right]\\
        &\hspace{6.2cm}\times t_\text{i}^{+\frac{1}{2}\sqrt{1-4\Omega_1}+\frac{1}{2}} t_\text{f}^{-\frac{1}{2}\sqrt{1-4\Omega_1}-\frac{1}{2}}\\
        &-\left[i\omega_{k}^{\text{I}} - \frac{1}{2t_\text{i}}\left(\sqrt{1-4\Omega_1}+1\right)\right]\left[i\omega_{k}^{\text{III}} - \frac{1}{2t_\text{f}}\left(\sqrt{1-4\Omega_1}-1\right)\right]\\
        &\hspace{6.2cm}\times t_\text{i}^{-\frac{1}{2}\sqrt{1-4\Omega_1}+\frac{1}{2}} t_\text{f}^{+\frac{1}{2}\sqrt{1-4\Omega_1}-\frac{1}{2}}\Bigg\},
    \end{split}
\end{equation}
and
\begin{equation}
    \begin{split}
        B_{k}^{\text{III}} = \frac{1}{Q\sqrt{2 \omega_k^{\text{I}}}} \frac{e^{-i(\omega^{\text{III}}_{k}t_\text{f} + \omega^{\text{I}}_{k}t_\text{i})}}{2i\omega^{\text{III}}_{k}\sqrt{1-4\Omega_1}} &\Bigg\{\left[i\omega_{k}^{\text{I}} + \frac{1}{2t_\text{i}}\left(\sqrt{1-4\Omega_1}-1\right)\right]\left[i\omega_{k}^{\text{III}} - \frac{1}{2t_\text{f}}\left(\sqrt{1-4\Omega_1}+1\right)\right] \\
        &\hspace{6.2cm}\times t_\text{i}^{\frac{1}{2}\sqrt{1-4\Omega_1}+\frac{1}{2}} t_\text{f}^{-\frac{1}{2}\sqrt{1-4\Omega_1}-\frac{1}{2}}\\
        & -\left[i\omega_{k}^{\text{I}} - \frac{1}{2t_\text{i}}\left(\sqrt{1-4\Omega_1}+1\right)\right]\left[i\omega_{k}^{\text{III}} + \frac{1}{2t_\text{f}}\left(\sqrt{1-4\Omega_1}-1\right)\right] \\
        &\hspace{6.2cm}\times t_\text{i}^{-\frac{1}{2}\sqrt{1-4\Omega_1}+\frac{1}{2}} t_\text{f}^{\frac{1}{2}\sqrt{1-4\Omega_1}-\frac{1}{2}}\Bigg\}.
    \end{split}
\end{equation}
\end{widetext}
We can then write \eqref{eq:planeWaveRegionIIIOpen} as in equation \eqref{eq:FLRWModeFunctionIIIvu}, in terms of the positive frequency plane waves in region III \eqref{eq:ModeFunctionUIII}. 

\subsection{Power-law expansion with exponent $\gamma = 3/2$}

At last, we consider $\gamma = 3/2$ with the scale factor in region II being
\begin{equation}
    a(t) = Q\abs{t}^{3/2}.
\end{equation}

\paragraph*{Region I.} We again omit this case as it is analogous to the discussion in Sec.\ \ref{subsec:powerLawgamma12}.

\paragraph*{Region II.} For $\gamma=3/2$, solving the mode equation yields
\begin{equation}
\begin{split}
    v_{k}^{\text{II}}(t) &= \frac{1}{t}\left[A_{k}^{\text{II}} J_2\left(2\sqrt{\frac{-h(k)}{Q^2 t}}\right)+ B_{k}^{\text{II}} Y_2\left(2\sqrt{\frac{-h(k)}{Q^2 t}}\right)\right].
\label{eq:ModeFunctionRegionIIOpenG15}
\end{split}
\end{equation}
The coefficients $A_{k}^{\text{II}}$ and $B_{k}^{\text{II}}$ can be computed from the matching condition in Eq.\ \eqref{eq:ModeFunctionsJointBoundaryi}, which gives
\begin{align}
    \begin{split}
        A_{k}^{\text{II}} &= \frac{\pi e^{-i \omega^{\text{I}}_k t_\text{i}}}{Q \sqrt{2 \omega^{\text{I}}_k t_\text{i}}} \times\Bigg[ - i \omega^{\text{I}}_k t_\text{i} Y_2\left(2\sqrt{\frac{-h(k)}{Q^2 t_i}}\right) \\
        &\hspace{2.2cm}+ \sqrt{\frac{-h(k) }{Q^2 t_\text{i}}}Y_1\left(2\sqrt{\frac{-h(k)}{Q^2 t_\text{i}}}\right)\Bigg]
    \end{split}
    \label{eq:CoefficientsRegionIIProperPhaseGamma15A}
\end{align}
and
\begin{align}
    \begin{split}
        B_{k}^{\text{{II}}} &= \frac{\pi e^{-i \omega^{\text{I}}_k t_\text{i}}}{Q \sqrt{2 \omega^{\text{I}}_k t_\text{i}}} \times\Bigg[ i \omega^{\text{I}}_k t_\text{i} J_2\left(2\sqrt{\frac{-h(k)}{Q^2 t_\text{i}}}\right) \\
        &\hspace{2.2cm}- \sqrt{\frac{-h(k) }{Q^2 t_\text{i}}}J_1\left(2\sqrt{\frac{-h(k)}{Q^2 t_\text{i}}}\right)\Bigg].
    \end{split}
    \label{eq:CoefficientsRegionIIProperPhaseGamma15B}
\end{align}

\paragraph*{Region III.} Again, we have the general solution
\begin{equation}
    v_{k}^{\text{III}}(t) = A_{k}^{\text{III}} e^{-i\omega_{k}^{\text{III}}t} + B_{k}^{\text{III}} e^{i\omega_{k}^{\text{III}}t},
\label{eq:planeWaveRegionIIIOpenG15}
\end{equation}
with frequency
\begin{equation}
    \omega_{k}^{\text{III}} =  \frac{\sqrt{-h(k)}}{a_\text{f}} = \frac{\sqrt{-h(k)}}{Qt_\text{f}^{3/2}}.    
\end{equation}
After matching at the boundaries, with Eq.\ \eqref{eq:BoundaryConditionf}, one finds
\begin{widetext}
\begin{equation}
    \begin{split}
        A_{k}^{\text{III}} &=  -\frac{\pi}{2}\frac{\sqrt{\omega_{k}^{\text{I}}t_\text{i}}}{Q t_\text{f}\sqrt{2}}  e^{-i\left(\omega_{k}^{\text{III}}t_\text{f} + \omega_{k}^{\text{I}}t_\text{i}\right)}\\
        &\hspace{1cm}\times \Bigg\{\Bigg[ iJ_1\left(2\sqrt{\frac{-h(k)}{Q^2 t_\text{f}}}\right) -  J_2\left(2\sqrt{\frac{-h(k)}{Q^2 t_\text{f}}}\right)\Bigg]\cdot \Bigg[ Y_1\left(2\sqrt{\frac{-h(k)}{Q^2 t_\text{i}}}\right) - i Y_2\left(2\sqrt{\frac{-h(k)}{Q^2 t_\text{i}}}\right)\Bigg]\\
        &\hspace{1.2cm}-\Bigg[ iY_1\left(2\sqrt{\frac{-h(k)}{Q^2 t_\text{f}}}\right) - Y_2\left(2\sqrt{\frac{-h(k)}{Q^2 t_\text{f}}}\right)\Bigg]\cdot \Bigg[ J_1\left(2\sqrt{\frac{-h(k)}{Q^2 t_\text{i}}}\right) - i J_2\left(2\sqrt{\frac{-h(k)}{Q^2 t_\text{i}}}\right)\Bigg]\Bigg\},
    \end{split}
\end{equation}
and
\begin{equation}
    \begin{split}
        B_{k}^{\text{III}} &=   \frac{\pi}{2}\frac{\sqrt{\omega_{k}^{\text{I}}t_\text{i}}}{Q t_\text{f}\sqrt{2}}  e^{i\left(\omega_{k}^{\text{III}}t_\text{f} - \omega_{k}^{\text{I}}t_\text{i}\right)}\\
        &\hspace{1cm}\times \Bigg\{\Bigg[ iJ_1\left(2\sqrt{\frac{-h(k)}{Q^2 t_\text{f}}}\right) +   J_2\left(2\sqrt{\frac{-h(k)}{Q^2 t_\text{f}}}\right)\Bigg]\cdot \Bigg[ Y_1\left(2\sqrt{\frac{-h(k)}{Q^2 t_\text{i}}}\right) - i Y_2\left(2\sqrt{\frac{-h(k)}{Q^2 t_\text{i}}}\right)\Bigg]\\
        &\hspace{1.2cm}-\Bigg[ iY_1\left(2\sqrt{\frac{-h(k)}{Q^2 t_\text{f}}}\right) +  Y_2\left(2\sqrt{\frac{-h(k)}{Q^2 t_\text{f}}}\right)\Bigg]\cdot \Bigg[ J_1\left(2\sqrt{\frac{-h(k)}{Q^2 t_\text{i}}}\right) - i J_2\left(2\sqrt{\frac{-h(k)}{Q^2 t_\text{i}}}\right)\Bigg]\Bigg\}.
    \end{split}
\end{equation}
\end{widetext}
The Bogoliubov coefficients $\alpha_k$ and $\beta_k$ follow from writing \eqref{eq:planeWaveRegionIIIOpenG15} as in equation \eqref{eq:FLRWModeFunctionIIIvu} in terms of the positive frequency plane waves in region III \eqref{eq:ModeFunctionUIII}.\\

\bibliography{references.bib}

\end{document}